\documentclass[a4paper,fleqn,usenatbib]{mnras}

\usepackage{newtxtext,newtxmath}

\usepackage[T1]{fontenc}
\usepackage{ae,aecompl}
\usepackage{subfig}
\usepackage{graphicx}	
\usepackage{amsmath}	
\usepackage{amssymb}	

\newcommand{\br}{{\mathbf r}}
\newcommand{\bv}{{\mathbf v}}

\usepackage[dvipsnames]{xcolor}
\def\alv#1{{\color{black}#1}}
\def\dch#1{{\color{black}#1}}
\def\pgb#1{{\color{black}#1}}

\title[Isolated models with primordial anisotropy]{\alv{The Kinematic Richness of Star Clusters -- I. Isolated Spherical Models with Primordial Anisotropy}}

\author[Breen, Varri \& Heggie]{
Philip G. Breen$^{1}$\thanks{E-mail: phil.breen@ed.ac.uk},
Anna Lisa Varri$^{2}$, 
Douglas C. Heggie$^{1}$
\\
$^{1}$School of Mathematics and Maxwell Institute for Mathematical Sciences, University of Edinburgh, Kings Buildings, Edinburgh EH9 3\dch{FD, UK}\\
$^{2}$Institute for Astronomy, University of Edinburgh, Royal Observatory, Blackford Hill, Edinburgh EH9 3HJ, UK}

\date{Accepted XXX. Received YYY; in original form ZZZ}

\pubyear{2017}

\begin{document}
\label{firstpage}
\pagerange{\pageref{firstpage}--\pageref{lastpage}}

\maketitle

\begin{abstract}

We investigate the dynamical evolution of isolated equal-mass star cluster models by means of direct N-body simulations, primarily focusing on the effects of the presence of primordial anisotropy in the velocity space. We found evidence of the existence of a {monotonic relationship} between the moment of core collapse and the amount and flavour of anisotropy in the stellar system. Specifically, equilibria characterised by the same initial structural properties (Plummer density profile) and with different degrees of tangentially-biased (radially-biased) anisotropy, reach core collapse earlier (later) than isotropic models. We interpret this result in light of an accelerated (delayed) phase of the early evolution of collisional stellar systems (``anisotropic-response"), which we have characterised both in terms of the evolution of the velocity moments and of a fluid model of two-body relaxation. For the case of the most tangentially anisotropic model the initial phase of evolution involves a catastrophic collapse of the inner part of the system which continues until an isotropic velocity distribution is reached. This study represents a first step towards a comprehensive investigation of the role played by kinematic richness in the long-term dynamical evolution of collisional systems.

\end{abstract}




\section{Introduction}

\alv{ The study of the {structure and} dynamical evolution of collisional stellar systems is often pursued under a relatively stringent set of simplifying assumptions, such as isotropy in the velocity space and the absence of order{ed} motions {(e.g.  rotation)}, which limit significantly the opportunity to explore the kinematic richness of star clusters. As a zeroth-order picture, two-body relaxation will certainly bring any dense, collisional system towards a state of approximate thermodynamical equilibrium, which, in phase space, may be characterised in terms of a quasi-Maxwellian distribution function. None the less, the often unexplored kinematic complexity may offer a refreshingly new and fertile degree of freedom which will enrich our fundamental understanding of the formation and dynamical evolution \dch{of} this class of stellar systems.  This is precisely the goal of the present study, which will be devoted to an exploration of the role of primordial (i.e., attributed to the initial conditions) anisotropy in the velocity space in the dynamical evolution of star clusters.  }
 
\alv{In this respect, possible deviations from isotropy in the velocity space may be explored along two complementary lines of investigations. On the one hand, after many decades of progressively more realistic numerical simulations, it is well
known that the dynamical evolution of collisional stellar systems, as driven by internal and external processes, may significantly affect the properties of their three-dimensional velocity space, generating, for instance, a variable degree of ``evolutionary'' anisotropy {(e.g., see \citealt{BM2003} and other references mentioned in the next paragraph)}. On the other hand, we might also ask whether, especially for star clusters characterised by long relaxation times, any signature of their formation process may actually be preserved in phase space, in the form of ``primordial'' anisotropy (e.g., see \citealt{Vesperini2014} and the discussion in the second part of this Section).}

\alv{In the first case, pioneering numerical experiments have recognised that anisotropy is indeed a natural outcome of star cluster dynamical evolution, especially when the system is in isolation \citep{Henon1971,SS1972}. Spitzer and collaborators have show\dch{n} that isolated globular clusters during their evolution develop a structure composed by two distinct regions: an isotropic core, and a radially anisotropic halo of stars, resulting from the scattering of stars from the centre, preferentially on radial orbits. Idealised models of isolated star clusters based on gaseous methods and linear perturbation theory have later confirmed that isolated systems tend to become progressively more anisotropic in their outer regions \citep[see][]{BS1986,S1991,LS1991}. Such a behaviour has been noted also by \citet{GH1994} in the context of the early exploration of the statistics of N-body simulations and by \citet{T1995} in two-integral Fokker-Planck models.} 

\alv{This picture has been subsequently extended to the inclusion in the models of the presence of an external tidal field, the effect of which is typically to curb the degree of radial anisotropy developed, most likely as a result of a preferential loss of stars on radial orbits and the general mass loss, which progressively exposes deeper and therefore more isotropic shells of the stellar system \citep{GH1997}. {Nevertheless,} the anisotropy profile remains radially-biased at intermediate radii, while it becomes isotropic \citep[or even mildly tangential, as illustrated by][]{T1997, BM2003} in the outer regions. Given these two limiting cases, it should not come as a surprise that the degree of  anisotropy developed in a collisional system strongly depends on the strength of the tidal field in which it is evolving \citep{TVV2016}, with a crucial role played by the population of potential escapers residing in the system \citep{CGZ2017}.}

\alv{On the side of distribution function-based models, these ideas have inspired the construction of a variety of equilibria, often defined as a direct generalisation of lowered isothermal models, in which the second integral of the motion is considered to be the specific angular momentum, either introduced via a simple exponential dependence (see \citealt{Michie1963}, \citealt{Davoust1977}, and, more recently, \citealt{GZ2015}), or by more complex prescriptions (e.g., see \citealt{Merritt1985},  \citealt{RT1984}, \citealt{Dejonghe1987}, \citealt{Dehnen1993}). Some of these models have been successfully applied to the interpretation of both observational data (e.g., see \citealt{GG1979}) and numerical simulations \citep{Sollima2015,Zocchi2016}.}

\alv{As for the investigation of ``primordial'' anisotropy, i.e. any deviation from velocity isotropy associated with the formation and early dynamical evolution phase, it might very well be that, especially for particularly rich and initially underfilling star clusters, their outer structure is not too far from that of bright elliptical galaxies for which violent relaxation is thought to have acted primarily to make the inner system quasi-relaxed, while the outer parts are only partially relaxed and therefore progressively more dominated by radially-biased anisotropic pressure (e.g., see \citealt{vanAlbada1982}). Some numerical experiments of violent relaxation have also been conducted in the presence of an external tidal field \citep{Vesperini2014} and it has been show\dch{n} that the configurations emerging at the end of such a cold collapse phase are characterised by a distinctive radial variation of the velocity anisotropy and can acquire significant internal differential rotation.} 

\alv{This line of argument motivated the construction of several families of dynamical models, to represent the final state of numerical simulations of the violent relaxation process (e.g., see \citealt{SB1985}; \citealt{Trenti2005}).
Some of these models have been successfully applied to the study of a sample of Galactic globular clusters under
different relaxation conditions \citep{Zocchi2012}, and recently modified to include an appropriate truncation in phase space, to heuristically mimic the effects of the external tidal field \citep{deVita2016}.}



\alv{We emphasise that, both in the case of evolutionary and primordial signatures, most of the attention has been devoted to the case of radially-biased anisotropy, also because of the interest in the regime in which collisionless equilibria may become unstable to the formation of a central bar, due to radial orbit instability (e.g., see \citealt{Poly1981,Fridman1984,Poly2015}). In turn, the corresponding regime of strong tangential anisotropy, and possible dynamical instabilities associated with it, has been only marginally explored (e.g, see \citealt{BHG1986, Weinberg1991}), partly because equilibria with tangential anisotropy are rare \citep[e.g., see][]{AE2006}. }

\alv{The exploration of the effects of kinematic complexity on the dynamical evolution of star clusters, especially in the form of deviations from velocity isotropy, is a particularly timely endeavour in light of two main reasons. First, from the observational point of view, it is now finally possible to access the full phase space of selected Galactic globular clusters and characterise their three-dimensional velocity space in terms of anisotropy profile and velocity maps, thanks to the synergy between the astrometric measurements conducted in the central regions of several globulars with Hubble Space Telescope \citep{Watkins2015} and those which will soon available from Gaia, especially in their periphery \citep{Pancino2017}.} 

\alv{Second, from the theoretical point of view, it is of paramount importance to fill the gap between the highly complex end state predicted by numerical simulations of star formation in a clustered environment and the extremely simplified initial conditions that are usually adopted to study the subsequent long-term dynamical evolution of star cluster\dch{s}. In this respect, some recent hydrodynamical simulations of the formation of you\dch{ng} massive star clusters have emphasised the importance of considering the kinematic dimension of protoclusters (e.g., see especially \citealp{LH2016} and \citealt{Mapelli2017}), also in view of the surprising morphological and kinematic richness that has emerged over the past few years in observational studies of young and intermediate age star clusters \citep[e.g., see][]{HB2012,Kuhn2015,Vincente2016}.}

\alv{In addition, a deeper understanding of the kinematic evolution of star clusters represents an essential step towards a satisfactory solution of topical open problems regarding the phase space characterisation of their elusive multiple stellar populations (see \citealt{Richer2013} and \citealt{Bellini2014} for the first observational assessments of the difference in velocity anisotropy between different stellar populations in Galactic globulars) and the existence of their putative intermediate mass black holes (see \citealt{Zocchi2017} for a recent study of the effect of velocity anisotropy of star clusters on the dynamical mass estimate of a central compact object). }

\alv{Before approaching the role played by this emerging kinematic complexity in any realistic setting, it is imperative to explore its effects on the gravothermodynamics of simple collisional stellar systems. For this reason, here we start from the case of spherical isolated systems and we consider a series of anisotropic equilibria which are characterised by the same spatial distribution of mass, but different properties of  their three-dimensional velocity space. This point is indeed essential, as it has been show\dch{n} that the structural properties of a stellar system do shape its subsequent collisional evolution (e.g., see \citealt{Quinlan1996}), especially the moment of core collapse, which will be the primary focus of our investigation (for an earlier study of the evolution of radially anisotropic systems towards core collapse, see \citealt{MPV1998}). The second, complementary part of this investigation will be devoted to the study of the role of {primordial rotation} (Breen, Varri, Heggie, in preparation).  }  

\alv{The paper is structured as follows. In Section~\ref{sec:df} we describe the distribution function-based models we use to define our initial conditions, characterised by spherical symmetry and anisotropy in the velocity space. In Section~\ref{sec:results} we present the results of a series of N-body simulations which have been performed to explore the role of anisotropy in the dynamical evolution of collisional systems in isolation. In Section \ref{sec:Dis} we outline the theory used to explain the early evolution of the N-body models and we discuss possible generalisations of our study. Finally, we present our conclusions in Section~\ref{sec:sum}. }

\begin{figure}
    \includegraphics[trim=0.7cm 1.cm 1.4cm 1.cm, width=\columnwidth]{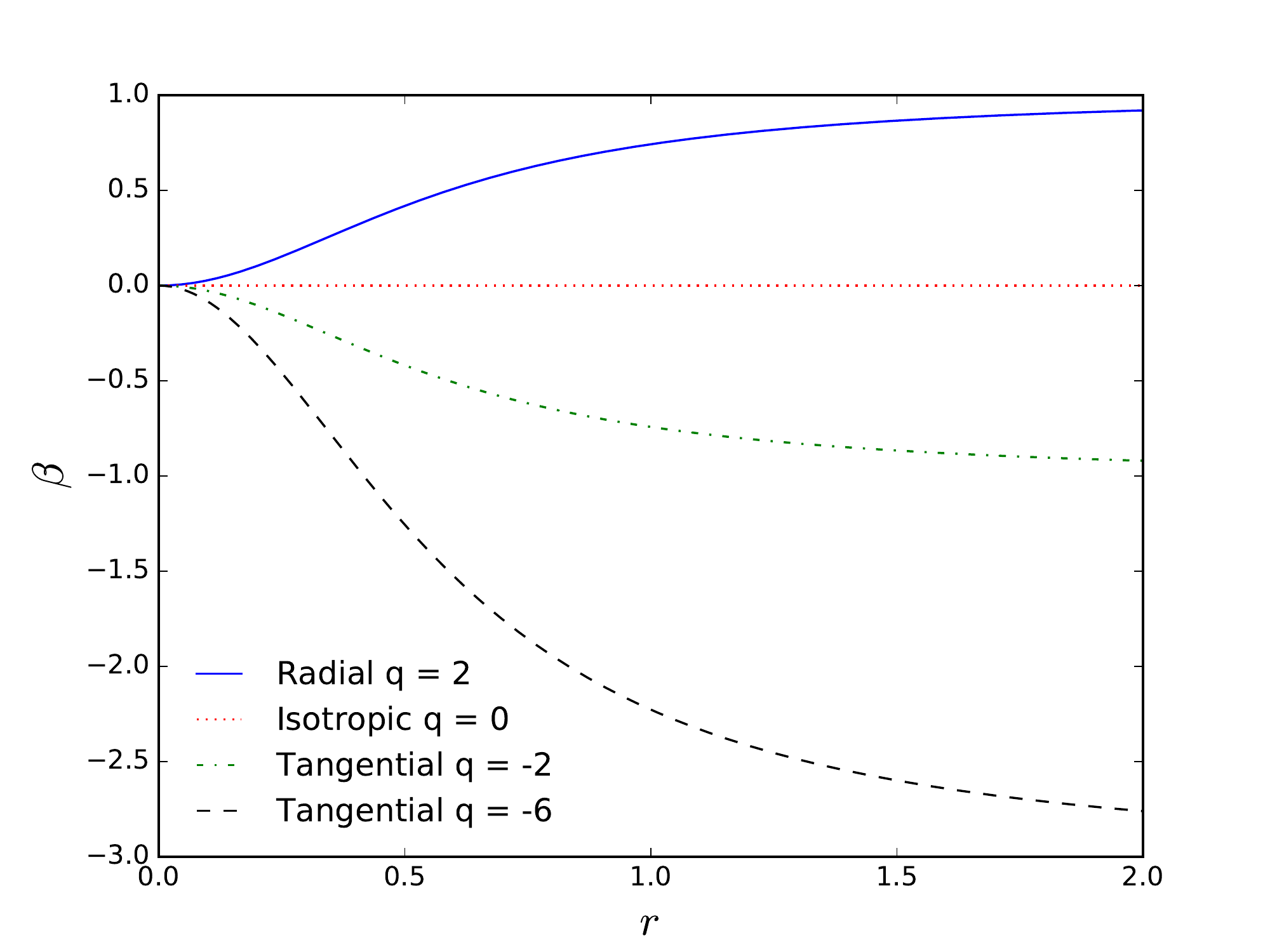}
    \caption{ \alv{Radial profiles of the anisotropy parameter $\beta (r)$ of four representative \alv{\citet{Dejonghe1987}} anisotropic Plummer models, corresponding to selected values of the main parameter $q$ (for definition, see Sect.~2.1). The half-mass radius is $r_h = (2^{2/3} -1)^{-1/2}  \approx 1.3$. The units are such that $G=M=a=1$.}  Note that curves with the same absolute value of $q$ are \alv{symmetric} by reflection across the x-axis, therefore the spatial distribution of their anisotropy is equivalent (with opposite sign).}
    \label{fig:aniso}
\end{figure}

\section{Method and initial conditions}
\label{sec:df}
{Unless otherwise stated, the quantities presented in all figures and table included in this paper are given in H\'{e}non units \citep{Henon1971}. A notable exception are the distribution functions defined in the present Section, \alv{in which a notation such that  $G=M=a=1$ has been adopted \citep{Dejonghe1987}}\dch{, where $M$ is the total mass and $a$ is the Plummer scale radius}. Where appropriate, we have also adopted the definition of the initial half-mass relaxation time ($t_{rh,i}$) as presented in equation (14.13) of \cite{HH03}.} 

\begin{figure}
\includegraphics[ trim=0.7cm 1.cm 1.4cm 1.cm, width=\columnwidth]{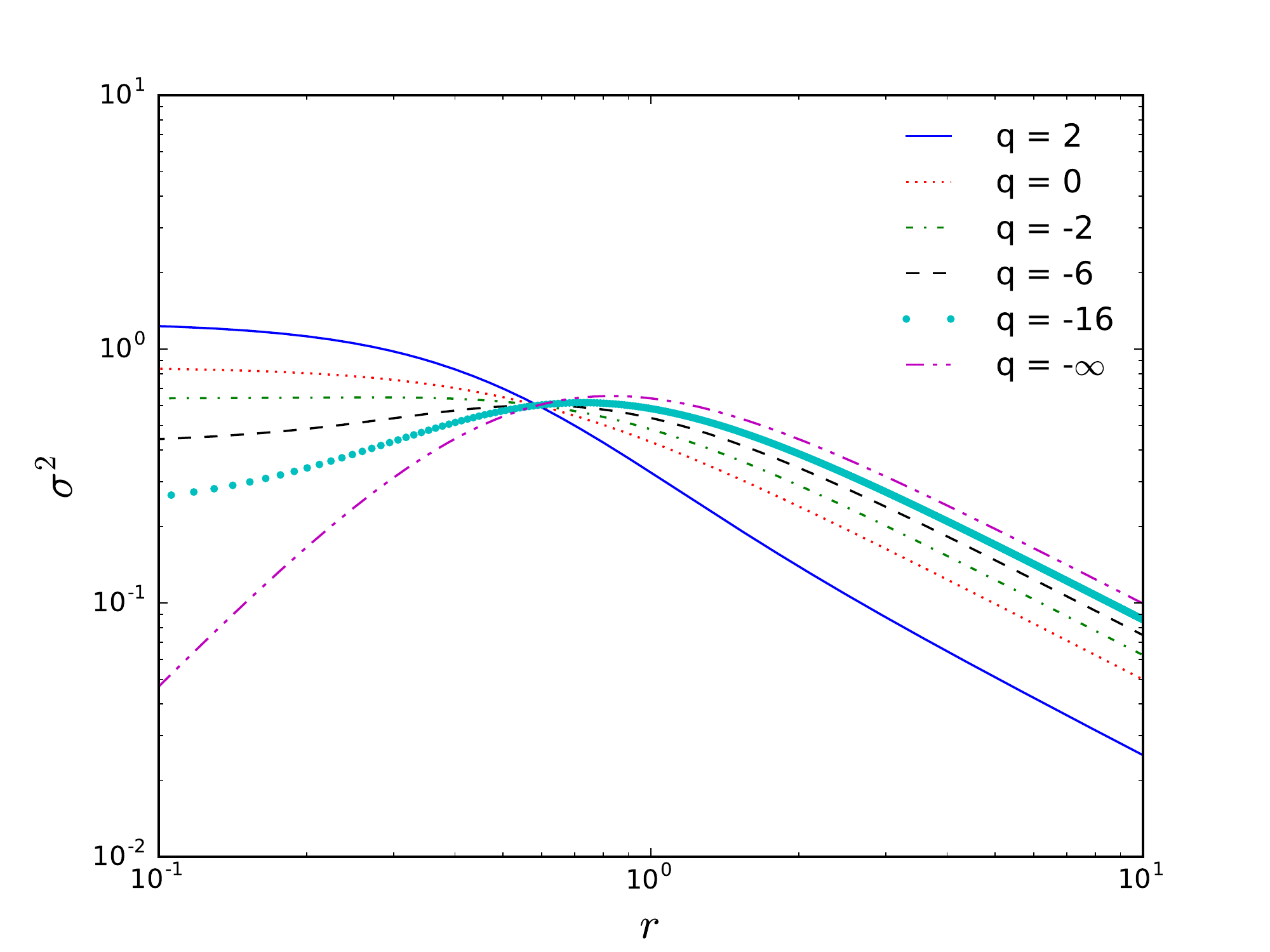}
\caption{\alv{Velocity dispersion profile \alv{of selected \citet{Dejonghe1987}} anisotropic Plummer models with $q =  -\infty, -16, -6, -2, 0$ and $2$. }The \alv{{inner} 
    gradient} \dch{increases} 
  with \alv{decreasing values of}  $q$ and becomes \dch{positive (i.e. with a temperature inversion)} 
  after $q = -\dch{3/}2$.
    \alv{The units system, {and the colour scheme and line style for the first four models,} are consistent with those adopted in Fig.~\ref{fig:aniso}.}}
    \label{fig:sig2}
\end{figure}

\subsection{Anisotropic Plummer models}
The initial conditions of the models presented in this study are realisations of \citet{Dejonghe1987} anisotropic Plummer models. The distribution function is given by: 
\begin{equation}
F_q(E,L) = \frac{3\Gamma(6-q)}{2(2\pi)^{\frac{5}{2}} \Gamma(q/2)} E^{\frac{7}{2}-q} H \left( 0, \frac{1}{2}q, \frac{9}{2}-q, 1;\frac{L^2}{2E} \right) 
\end{equation}
where $E$ is energy, $L$ angular momentum, $q$ is the parameter which controls the amount of anisotropy (and lies in the range $-\infty < q  < 2$), $\Gamma$ is the Gamma function and $H$ is a function, which \alv{may be expressed} in terms of the hypergeometric function ${}_2F_1$. For $ x \leq 1 $ \\
\begin{multline}
H(a, b, c, d; x) = \frac{\Gamma(a+b)}{\Gamma(c-a)\Gamma(a+d)} x^a\\  {}_2F_1(a+b, 1+a-c; a+d; x)  
\end{multline}
and for $ x > 1 $
\begin{multline}
H(a, b, c, d; x) =  \frac{\Gamma(a+b)}{\Gamma(d-b)\Gamma(b+c)} \frac{1}{x^{b}}\\ {}_2F_1\left(a+b, 1+b-d; b+c; \frac{1}{x}\right)\dch{.}
\end{multline}
\alv{Depending on the value of $q$, the resulting models are characterised by isotropy ($q=0$), tangential ($q < 0$) or radial ($q>0$) anisotropy. For these equilibria, the anisotropy parameter $\beta$ has the convenient form} 
\begin{equation}
\beta \alv{(r)} = 1 - \frac{\sigma_t^2}{2\sigma_r^2} = \frac{q}{2}\frac{r^2}{1+r^2} ,
\label{eqn:beta}
\end{equation}
    {where $\sigma_t$ and $\sigma_r$ are the \dch{rms} tangential \dch{(two-dimensional)} and radial \dch{velocity} components
, respectively.}
\alv{For reference,} the $\beta$ profiles for some of the models \alv{considered in this study are illustrated} in Fig. \ref{fig:aniso}.{ We note that models with different values of q are characterized by different energy distributions (for a complete description, see Dejonghe 1987, sections 3.3 and 3.4). Nevertheless the total energy and the half-mass relaxation time of all models are the same (in H\'{e}non units). }

\alv{Discrete numerical realisations of the selected models within the family described above were generated using a similar technique as in \citet{Aarseth1974}, adapted for the properties of the anisotropic distribution functions under consideration}\footnote{\pgb{In the isotropic case the \dch{magnitude of the} velocit\dch{y} ($v$) \dch{is} sampled from \dch{a} 
    one-dimensional distribution $\dch{\propto} f(r,v)v^2$, while here, in the anisotropic case, the \dch{magnitudes of the} tangential and radial velocities $(v_t,v_r)$ have been sampled from \dch{a} 
    joint probability distribution $\dch{\propto} f(r,v_t,v_r)v_t$.}  {For acceptance-rejection sampling a bound on the maximum value of the $f(r,v_t,v_r)v_t$ has to be known at each radius. This is calculated numerically in advance of sampling on a radial grid. During sampling the bound at a particular r is calculated by interpolation. }}.

\subsubsection{\alv{Radially-biased models}}\label{sec:OMmodels}
The most radially anisotropic model in the Dejonghe series is the $q=2$ model. This \alv{configuration} is the same as one of the Osipkov-Merritt models \citep{Osipkov1979,Merritt1985}\dch{, i.e. the one} with \dch{anisotropy radius} $r_a = 1$. The distribution function of the radially anisotropic Osipkov-Merritt Plummer model \alv{may be defined as}
\begin{equation}
f(Q) = \frac{\sqrt{2}}{378\pi^3G\sigma(0)} \Bigg ( \frac{-Q}{\sigma^2(0)} \Bigg )^{\frac{7}{2}} \Bigg [   1 - \frac{1}{r_a}  + \frac{63}{4r_a^2}   \Bigg ( \frac{-Q}{\sigma^2(0)} \Bigg )^{-2}  \Bigg ]   
\label{OM_DF}
\end{equation}
where $Q = E + J^2/(2r_a^2)$ {and $\sigma(0) $ is the central value of the (scalar\dch{, i.e. one-dimensional}) velocity dispersion
}. Here the amount of \alv{radial} anisotropy grows with decreasing $r_a$. \alv{The definition given in Equation \ref{OM_DF}} is valid for $r_a > 0.75$; \alv{for lower values of $r_a$} the distribution function becomes negative. \alv{In principle, the Osipkov-Merritt} models can be used to extend the Dejonghe series to \alv{configurations characterised by a higher level of radial anisotropy (i.e., beyond the limiting case of such series, as set by $q = 2$).} 

\alv{For the purpose of this study, we have considered the Dejonghe model with $q=2$ ($r_a=1$), and, in order to explore the stability threshold of these equilibria with respect to radial orbit instability, we have also explored the evolution of the Osipkov-Merritt configuration with $r_a=0.75$. A discussion of such threshold is presented in Section~\ref{sec:roi}.}


\subsubsection{\alv{Tangentially-biased models}}

\alv{Models with negative values of the parameter $q$ are characterised by tangential anisotropy; for the purpose of the current investigation we have considered the cases with $q=-2,-6,-16, -\infty$; all configurations ($N=8192$) appear to be dynamically stable. }

\pgb{ \cite{Merritt1985} also considered tangential anisotropic distribution functions based on the parameter $Q_{-} = E - J^2/(2r_a^2)$ (called ``Type II"). However, in order to obtain distributions functions which are easily \alv{invertible from the} density profile, \alv{only the portion of the domain such that $ r < r_a $ should be considered}. In order to extend the models beyond $r_a$, \cite{Merritt1985} offered two solutions: ``Type IIa" \alv{is such that the density distribution is completed} by adding circular orbits outside $r_a$ and ``Type IIb"  \alv{by adding orbits that correspond to the condition of} constant $Q_{-}$. 
Both methods result in a discontinuity in the tangential velocity dispersion at $r_a$ \citep[cf][]{Merritt1985}.  Due to the additional complications that arise for $r>r_a$ , \alv{in the present investigation} we will not consider tangential anisotropic Osipkov-Merritt models further.}

\begin{table}
\caption{\alv{Properties of the N-body simulations. From left to right, the columns report the values of the $q$ parameter of the \citet{Dejonghe1987} anisotropic Plummer model,  the number of particles $N$, the number of realisations performed $N_{r}$, the average core collapse time $t_{cc}$ (with the standard deviation, in units of $t_{rh,i}$), the difference in collapse time compared to \dch{the} isotropic model $\Delta t_{cc}$, \dch{and twice} the ratio of the kinetic energy of radial motions $T_r$  to the kinetic energy of transverse motions $T_\perp$. Note that the standard deviation quoted for model $q=-6$ is most likely underestimated due to low number statistics.} }
 \label{tab:tab1}
 \begin{tabular}{||c c c c c c ||} 
 \hline 
   $q$ \quad & $N$ & \quad $N_{r}$ \quad & \quad $t_{cc}$ \quad &   \quad $\Delta t_{cc}$ \quad & \quad  $2\,T_r/T_\perp$ \\ [1.0ex] 
 \hline\hline
  2 & 8192 & 4 & 19.05 $\pm$ 0.57 &  2.01 & 1.96  \\ 
 \hline
  0 & 8192 & 4 &17.04 $\pm$ 0.47 &  0.00 & 1.00 \\
 \hline
 -2 & 8192 & 4 & 14.99 $\pm$ 0.50 &  -2.05 & 0.66 \\
 \hline
 -6 & 8192 & 4 & 12.32 $\pm$ 0.17 &  -4.73 & 0.40 \\
 \hline
-16 & 8192 & 4 & 9.99 $\pm$ 0.62 &  -7.06 & 0.20 \\
 \hline
 -$\infty$ & 8192 & 4 &  6.16 $\pm$ 0.36 &  -10.88 & 0.00  \\ 
 \hline
\end{tabular}
\end{table}

\subsubsection{Einstein Sphere}\label{sec:einstein}

For $q \to -\infty $, \alv{the model representing the limiting case in the series of tangentially anisotropic Plummer equilibria consists exclusively of stars on circular orbits.} This configuration is referred to as ``Einstein Sphere'' \citep{Fridman1984} as it was first studied by \citet{Einstein1939} in the context of \alv{an exploration of a class of spherically-symmetric solutions of the field equation in General Relativity. Such a class represented a static cluster of collisionless gravitating particles, as a tool to investigate the physical significance of the Schwarzschild singularity.} It is straightforward to construct Einstein Spheres for any spherically symmetric density distribution. These models have the unusual feature that  $\sigma^2 \to 0$ as $r \to 0$.  As $q$ decreases {(in the Dejonghe 1987 sequence)}, 
the gradient of the velocity dispersion {at small radii} \dch{increases}, eventually becoming \dch{positive} 
(at $q = -\dch{3/}2$). \alv{The radial profiles of the velocity dispersion calculated for a selection of models are presented in Fig.~\ref{fig:sig2}.}

\alv{One may wonder if there is a radial analogue of the Einstein Sphere, i.e. a Plummer model consisting of entirely radial orbits. It can be shown that models with a finite central density can not be constructed by means exclusively of radial orbits \citep{BJ1968,RT1984}. }

\begin{figure}
	\includegraphics[trim=0.8cm 1.1cm 0.4cm 0.1cm, width=\columnwidth]{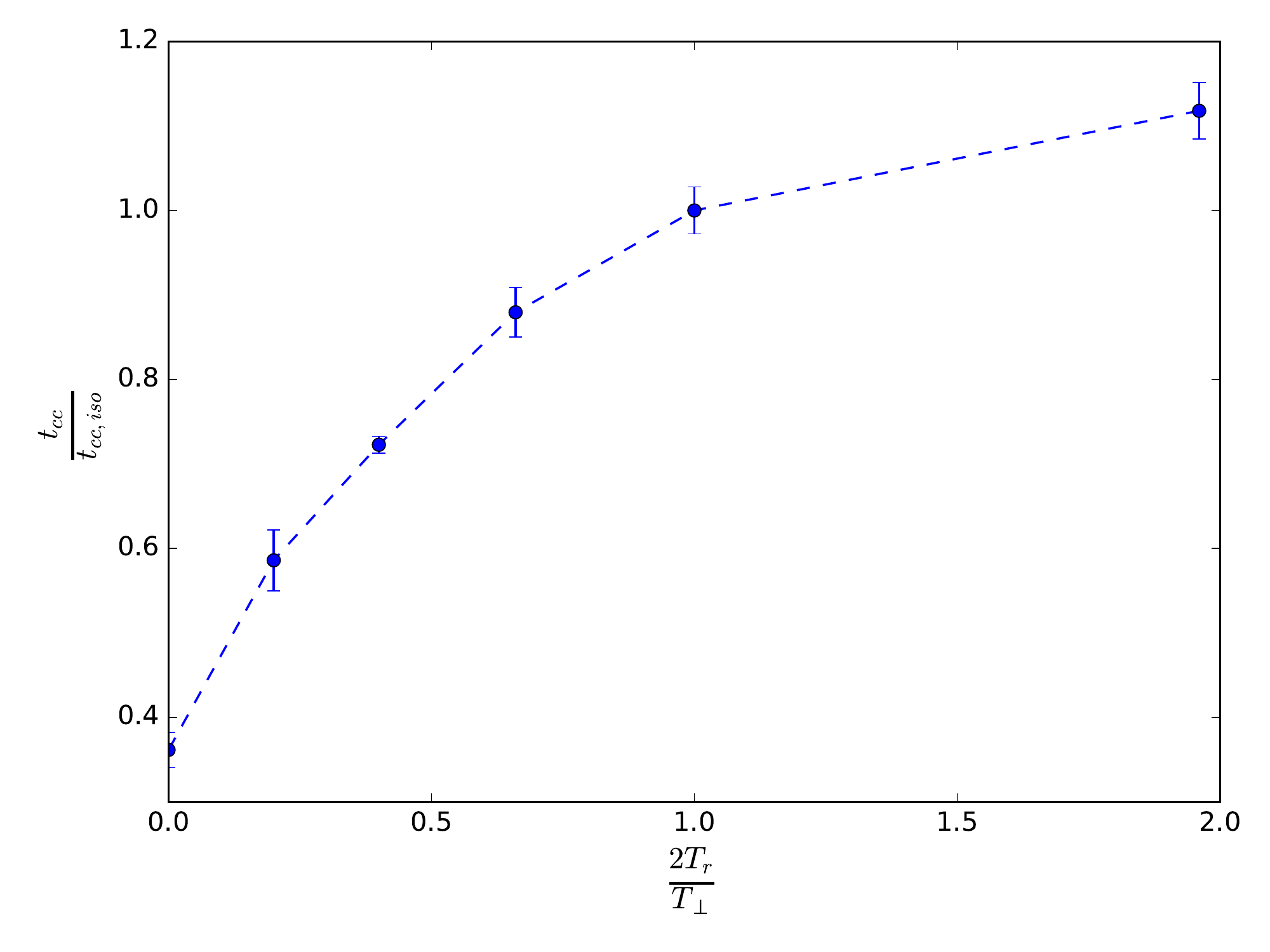}
     \caption{\alv{Time of core collapse of all models (normalised to the \dch{average} core collapse time of the isotropic Plummer model) $t_{cc}/t_{cc,iso}$ expressed as a function of the initial value of \dch{twice} the ratio of the radial to tangential kinetic energy $2\,T_r/T_{\perp}$. The error bars are calculated from the standard deviations presented in Table~\ref{tab:tab1}. A simple \dch{piecewise} linear interpolation is illustrated as a dashed line, to guide the eye. }}
   \label{fig:tcc}
\end{figure}

\subsection{\alv{Numerical simulations}}\label{sec:numerical}

\begin{figure}
	\includegraphics[trim=0.6cm 0.8cm 1.4cm 1.cm, width=\columnwidth]{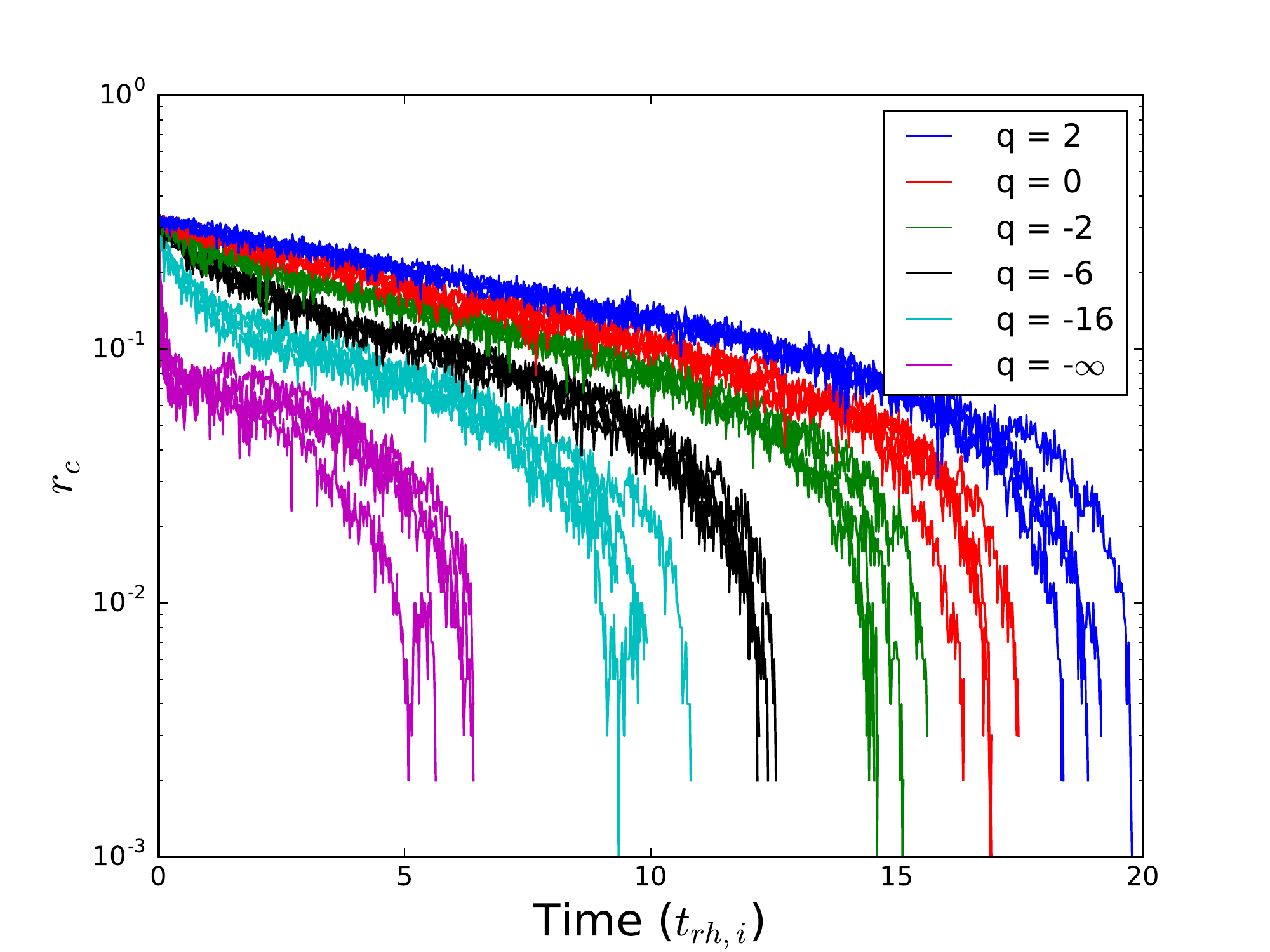}
    \caption{ \alv{Time evolution of the core radius of selected anisotropic Plummer models with $q=  -\infty, -16, -6, -2, 0 ,2$. Time is expressed in units of the initial half-mass relaxation time ($t_{rh,i} \approx 112 $ H\'{e}non units \alv{for all models}). For $q<0$ the models are tangentially anisotropic, for $q=0$ isotropic and for $q>0$ radially anisotropic (see Section~\ref{sec:df}). The four different realisations of each model are depicted with the same colour.{ The six groups of curves correspond to the six values of q, which increases from the group at the extreme left to the group at the extreme right.}}
    }
    \label{fig:main}
\end{figure}

\alv{Starting from the initial conditions described in the previous Section, we have performed a series of direct N-body simulations to explore the long-term evolution of the anisotropic equilibria under consideration. Each model has been studied by means of four independent numerical realisations\footnote{\dch{This refers to the Dejonghe models listed in Table \ref{tab:tab1}.   Only one realisation of the Osipkov-Merritt model with $r_a = 0.75$ was computed}}. \dch{All models used stars of equal mass with no primordial binaries and no stellar evolution.}  The properties of the simulations are presented in Table~\ref{tab:tab1}. The simulations were conducted by means of the direct summation code NBODY6}\footnote{\alv{NBODY6 is publicly available for download from
www.ast.cam.ac.uk/$\sim$sverre/web/pages/nbody.htm}.
NBODY6 comes with an option to generate an isotropic Plummer model. However, this option imposes a cut-off radius, \alv{which we have not adopted when generating our initial conditions.} The amount of mass outside the cut off is approximately 1.5\%. When the cut off is included it appears to increase the core collapse time by roughly 50 H\'{e}non time units.} \citep{Aarseth2003, Nit2012}, enabled for use with Graphical Processing Units (GPU). \alv{We have evolved the models in isolation, without any prescription for removing particles, until they all reached core collapse (as identified from the evolution of the core radius, see Fig.~\ref{fig:main}). }

\section{Results}
\label{sec:results} 

\subsection{{Dependence of core collapse on anisotropy}}
The core collapse times ($t_{cc}$) \alv{of the anisotropic models under consideration} are presented in Table~\ref{tab:tab1}. There is a clear dependence of the core collapse time on the amount of anisotropy in the system, \alv{ as expressed by} the parameter $q$ or $2\,T_r/T_{\perp}$. \alv{ Specifically, we found that the model characterised by radial anisotropy ($q=2$) requires a longer time to reach core collapse, compared to the {standard} 
  isotropic case; tangentially anisotropic models require a shorter time, with the most tangential case ($q=-\infty$, Einstein Sphere) being the absolute fastest to collapse. On the basis of the series of simulations we have performed, we therefore conclude that, in the presence of the same initial structural properties, {there} 
  appears to exist a {monotonic relationship} between the amount (and flavour) of primordial anisotropy and the \dch{time} 
  of core collapse for spherical, isolated collisional systems.} 

The relationship between $2\,T_r/T_{\perp}$  and the collapse time is shown in Fig~\ref{fig:tcc}. \alv{It is essential to emphasise that this relationship {is likely to} 
  depend on the initial structural properties of the models and the spatial distribution of their primordial anisotropy, {and the evidence we present for it} 
  is limited to the anisotropic Plummer models under consideration. None the less, it {suggests that} 
  one may certainly consider exploring such a two-dimensional parameter space also for other families of anisotropic equilibria. Some considerations regarding the effect of the spatial distribution of the anisotropy are presented in Section~\ref{sec:geniso}.}

The evolution of the core radius ($r_c$) of all the models is shown in Fig.~\ref{fig:main}. \alv{Different numerical realisations of the same anisotropic equilibrium are depicted with the same colo\dch{u}r, to offer a visual representation of the spread of the core collapse times (see also the standard deviations reported in Table~\ref{tab:tab1})}. The small spread in values for $q=-6$ is likely due to the small sample size \alv{(four realisations)}.

\begin{figure}
	\includegraphics[trim=0.6cm 1.cm 1.4cm 1.cm, width=\columnwidth]{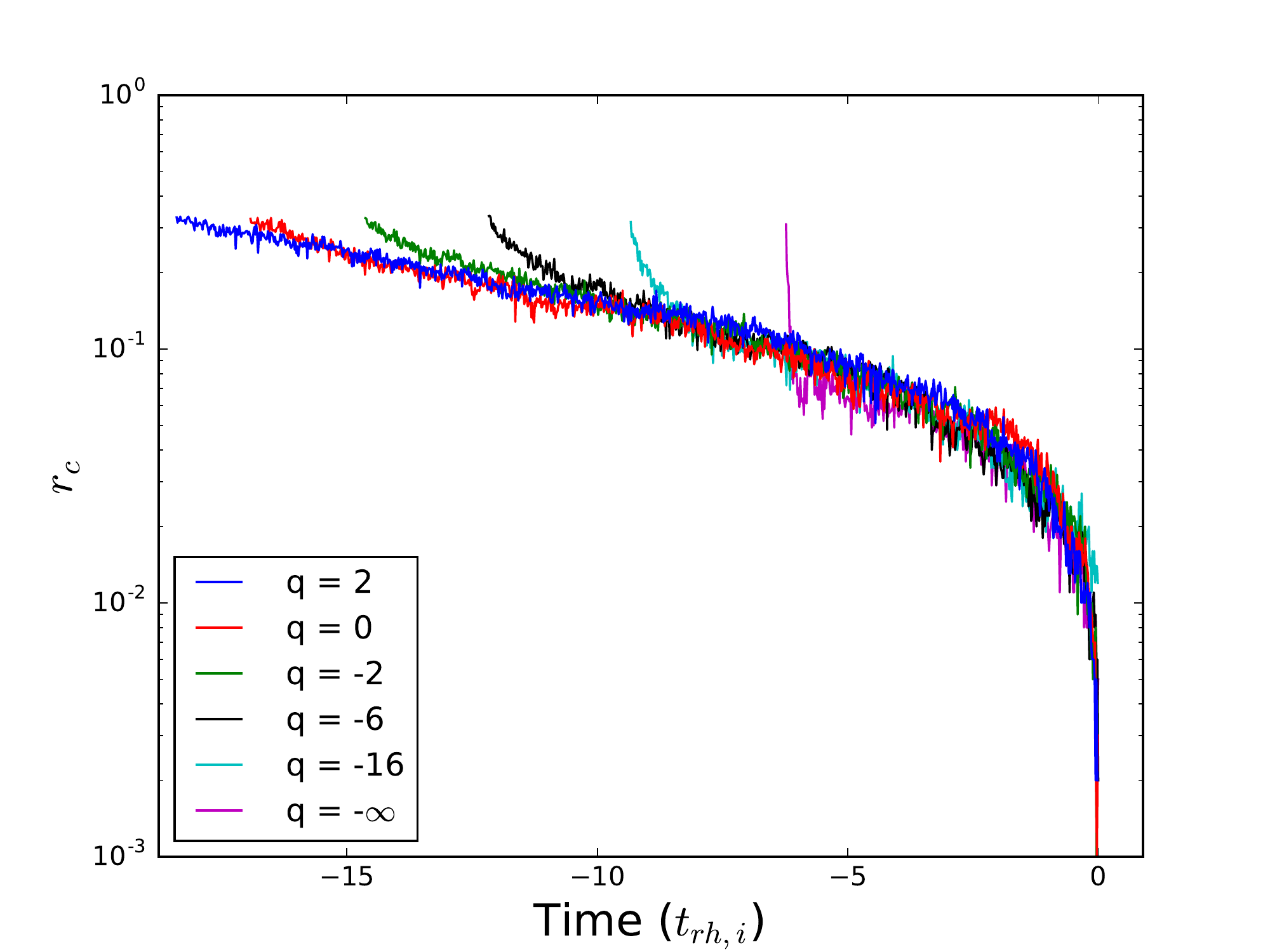}
        \caption{Time evolution of the 
          \alv{value of the core radius} of the anisotropic Plummer models illustrated in Fig.~\ref{fig:main}\dch{, though only one realization is shown for each value of q.}. \alv{The time variable is scaled} so that core collapse happens at time $t'=0$ (i.e. $t' = t - t_{cc} $). { The abscissa at the starting point of a curve is
a decreasing function of q.}  The evolution of $r_c$ becomes similar once the inner part of the system reaches a {nearly} isotropic velocity distribution. }
    \label{fig:main2}
\end{figure}

If we rescale \alv{the time variable} so that core collapse occurs at \alv{$t'=0$ (i.e. $t' = t - t_{cc} $)}, as \alv{performed} in Fig~\ref{fig:main2}, \alv{it may be noticed that} all the models evolve in a similar way once the inner part of the system becomes {approximately} isotropic \alv{(for an evaluation of the time evolution of the level of anisotropy in the models, see Fig.~\ref{fig:main3})}. 

\begin{figure}
	\includegraphics[trim=0.6cm 1.cm 1.4cm 1.cm, width=\columnwidth]{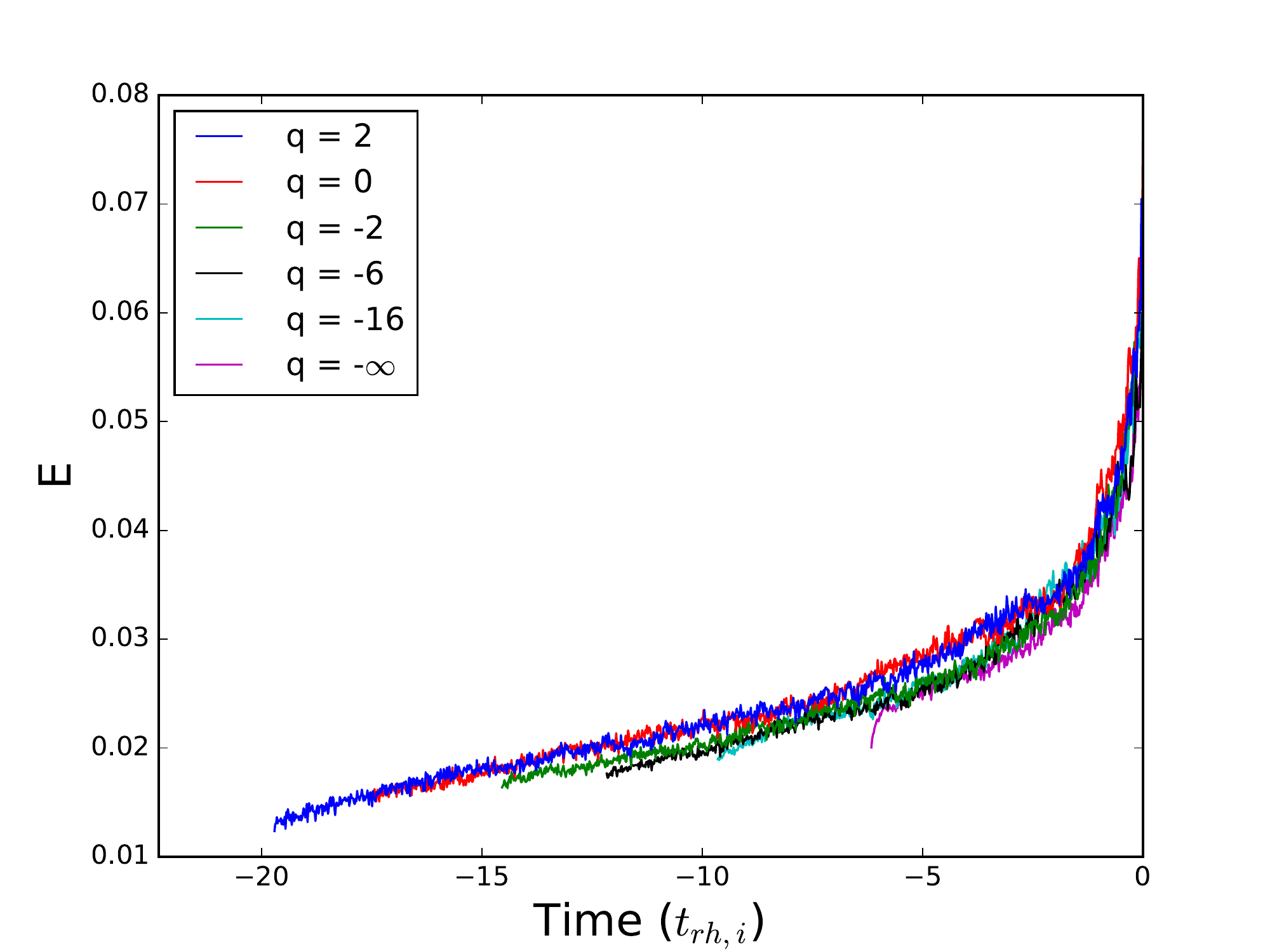}
    \caption{A measure of energy of the innermost 100 particles, where $E = \dch{m}\sum \phi(r_i) - 0.5v_i^2 $ and $\phi$ is calculated assuming spherical symmetry and using the same method as described by  \citet{Henon1971}. The models \dch{approach and then} follow a similar evolutionary track in energy space. \pgb{One realization is shown for each value of q.{ The curves are ordered as in Fig.5.}}}
    \label{fig:E}
\end{figure}

\alv{An assessment of the energy of the innermost } 100 particles is offered in Fig. \ref{fig:E}. \alv{In order to make the energy evaluation} more efficient, we calculated the potential by assuming spherical symmetry and using the method described by \citet{Henon1971}, then we simply sum\dch{med} the energy of \dch{the} innermost 100 particles (i.e. $E = \dch{m}\sum \phi(r_i) - 0.5v_i^2 $)\dch{, where $m$ is the individual stellar mass}. \alv{As in the case of the evolution of the core radius, once a condition of isotropy in the central regions is approached, the models tend to evolve on {nearly} the same track in the energy space.}

\alv{We wish to note} that none of the models listed in Table~\ref{tab:tab1} show signs of radial orbit instability, even though the model with $q=2$ \alv{has a value $2 T_r/T_\perp =1.96$}, \alv{which sits at} the upper limit of the critical range ($1.7 \pm 0.25$) found by \citet{Poly1981}. In Section~\ref{sec:roi} we extend this series to explore \alv{one example of the} more radially anisotropic Osipkov-Merritt configurations, \alv{which are indeed unstable. Our attention to the emergence of possible signs of radial orbit instability is motivated by the fact that such a process, by determining the presence of a central bar, would significantly modify the mass distribution of the system, therefore introducing a crucial element of structural difference which would affect the validity of our comparative analysis of the \dch{time} 
  of core collapse.}

\subsection{Early evolution: the anisotropic response }
\alv{For a fixed spatial distribution of mass,} as we change the amount of anisotropy \alv{in the system} we are also changing the velocity dispersion profile, \alv{as illustrated} in Fig.~\ref{fig:sig2}. The central velocity dispersion decreases
with decreasing $q$ and a temperature inversion develops below $q = -\dch{3/}2$.  In the case of gravothermal oscillations \citep[e.g. see][]{BS1984}, a temperature inversion is \alv{usually} associated with a phase of expansion. \alv{In such a case, it determines an evolutionary condition such that} the heat flows from the hotter outer regions to the cooler inner regions, \alv{causing the system to expand}. \alv{Interestingly, in the current case,} we observe the opposite behaviour, where the inner regions actually collapse faster with increasing negative temperature gradient. \alv{{Below} we {will} interpret this behaviour as a collapse determined by the ``conversion" of the tangentially-biased motions (as measured by $\sigma_t^2$) into radial ones ($\sigma_r^2$)} as the system relaxes towards an isotropic velocity distribution. 
For the case of radial models we predict the opposite effect. 

\begin{figure}
	\subfloat[]{{\includegraphics[trim=0.0cm 1.cm 1.4cm 0.7cm, width=\columnwidth]{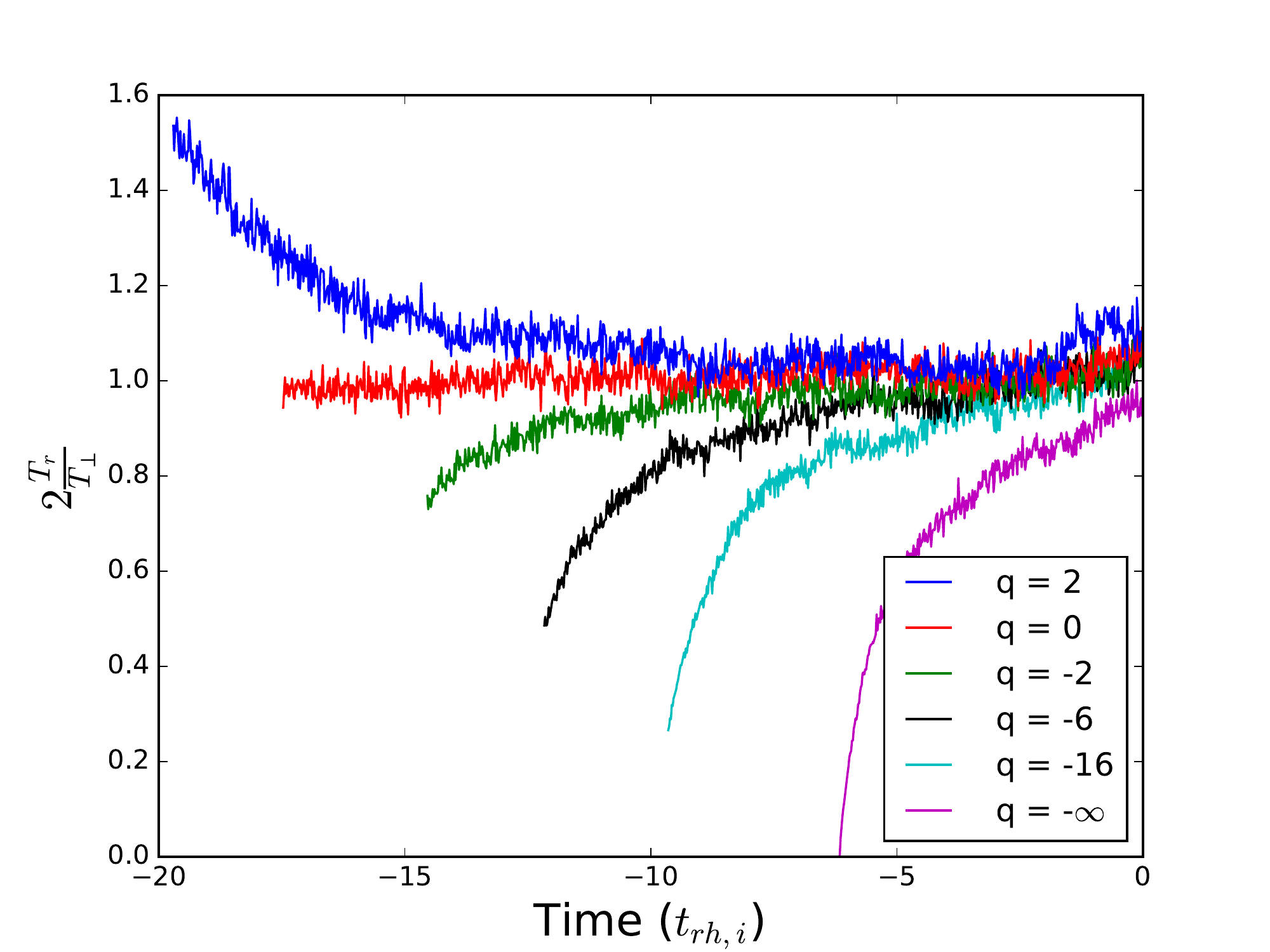} }}%
	\quad
	\subfloat[]{{\includegraphics[trim=0.0cm 1.cm 1.4cm 0.6cm, width=\columnwidth]{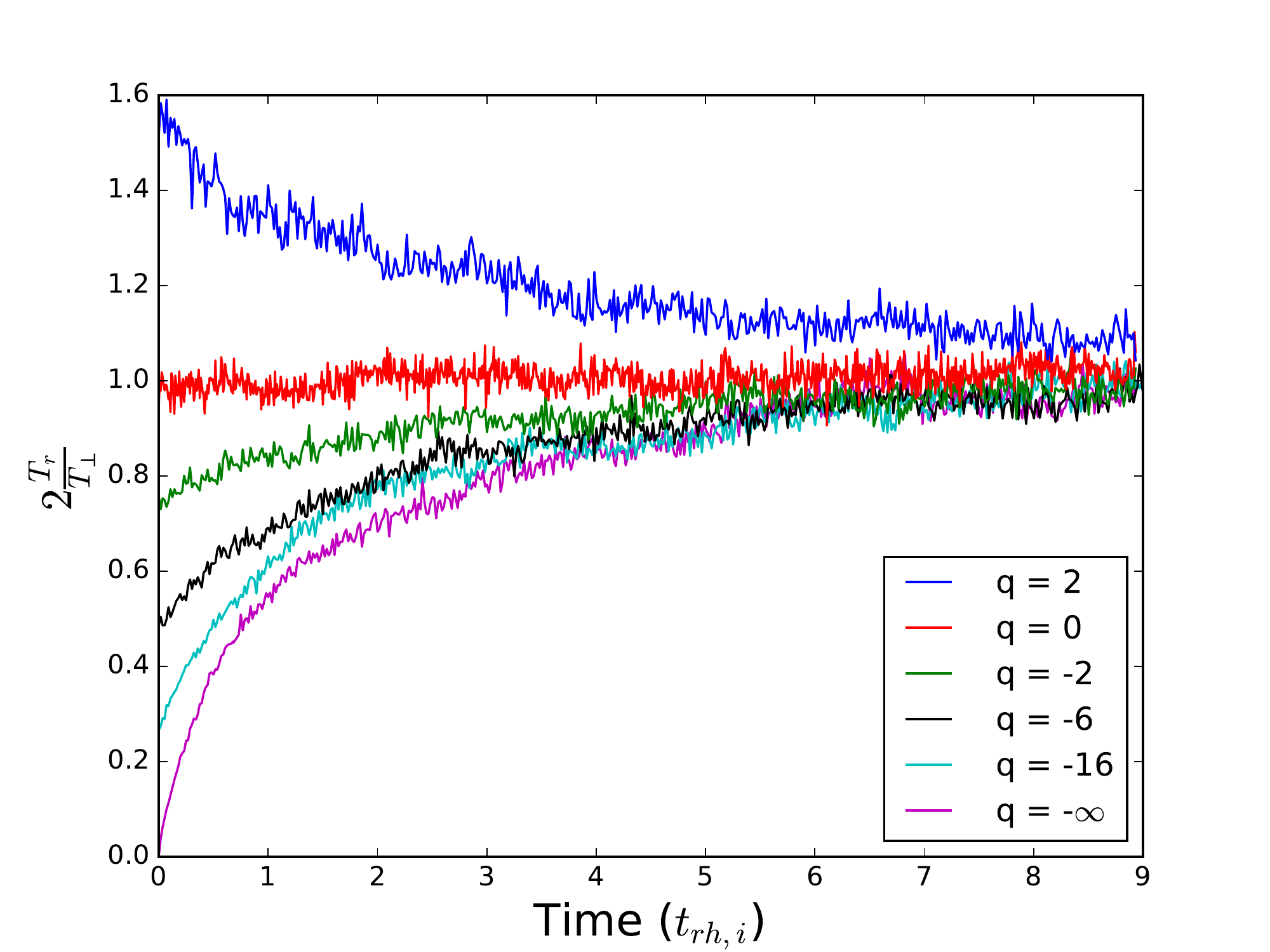} }}%
	\quad			
	\subfloat[]{{\includegraphics[trim=0.0cm 1.cm 1.4cm 0.6cm, width=\columnwidth]{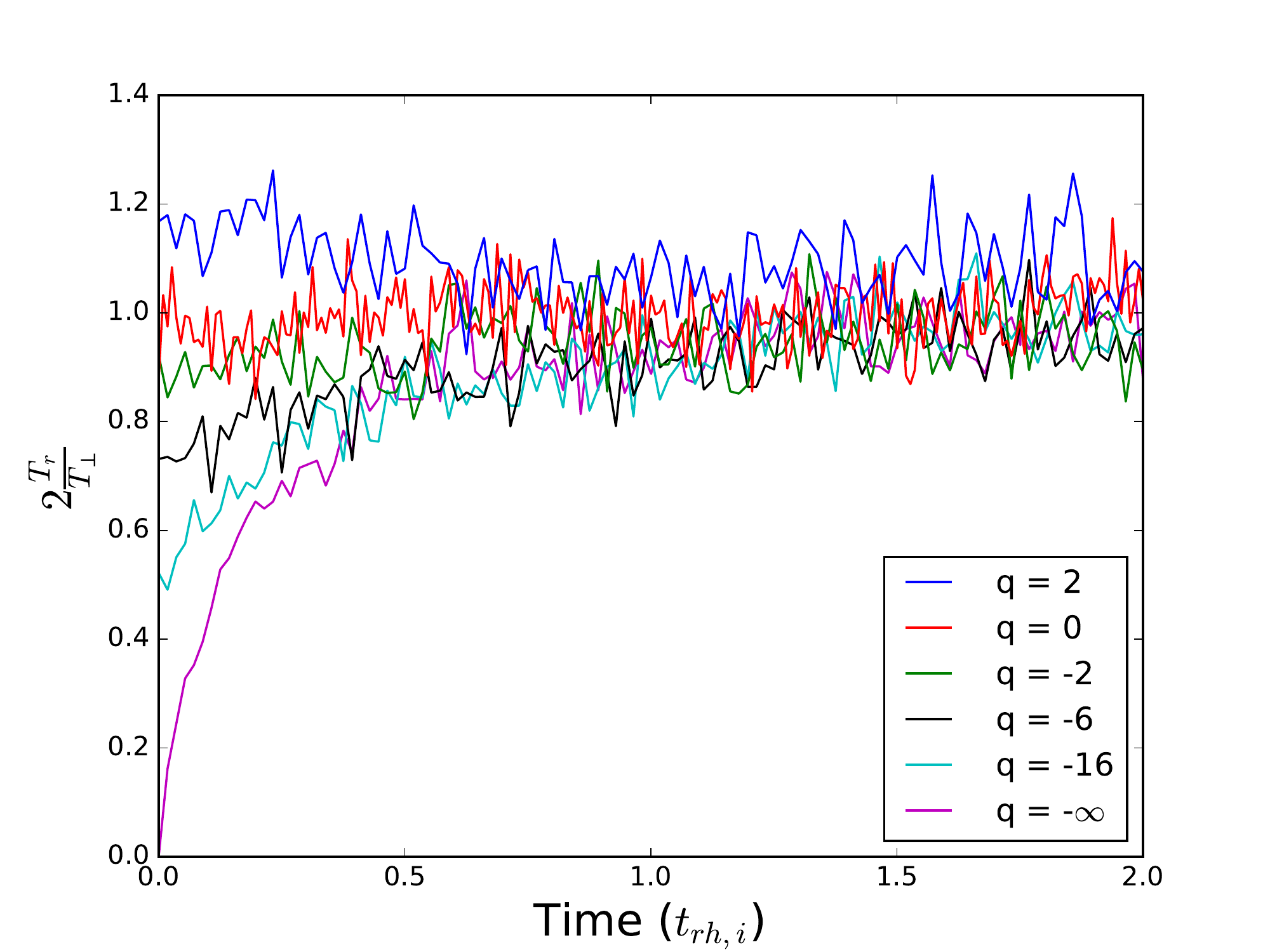} }}%
    \caption{\alv{Time evolution of the velocity anisotropy in the central regions o\dch{f} the models illustrated in Fig.~\ref{fig:main2}\dch{.} {From top to bottom, the curves are in order of decreasing q, except where they overlap}. \pgb{Top: $2\,{T_r}/{T_\perp}$ within the half-mass radius \dch{with the time} scaled so that $t_{cc}=0$. Middle: Same as Top without shifting time scale}. Bottom: $2\,{T_r}/{T_\perp}$  within the 10\%  Lagrangian radius. The velocity distribution of the mass within the 10\% and 50\% Lagrangian radii becomes \dch{approximately} isotropic \dch{for all models} after 50 and 700 H\'{e}non units, respectively ($t_{rh,i} \approx 112 $ H\'{e}non units \alv{for all models}). }}
\label{fig:main3}
\end{figure}

\alv{By increasing the initial amount of tangential anisotropy in the system}, there is a more pronounced collapse in the initial phase of evolution which can be clearly seen in Fig. \ref{fig:main}. As we will argue in Sections~\ref{sec:jeans} {and \ref{sec:theoryII}}, \alv{in which we provide a quantitative interpretation of this behaviour in terms of the evolution of the fluid moments}, we expect the inner parts of the system to \alv{lose support from the tangentially-biased motions} as the system relaxes towards an isotropic velocity distribution. For the most tangential case \alv{in our series} this process results in a catastrophic collapse, as the inner regions \alv{rapidly evolve towards} an isotropic velocity distribution. {The reason for this is the very short central relaxation time, which is caused by the low central velocity
dispersion (see Fig.~\ref{fig:sig2}).}

\begin{figure}
	\subfloat[]{{\includegraphics[trim=0.6cm 1cm 1.4cm 0.7cm, width=\columnwidth]{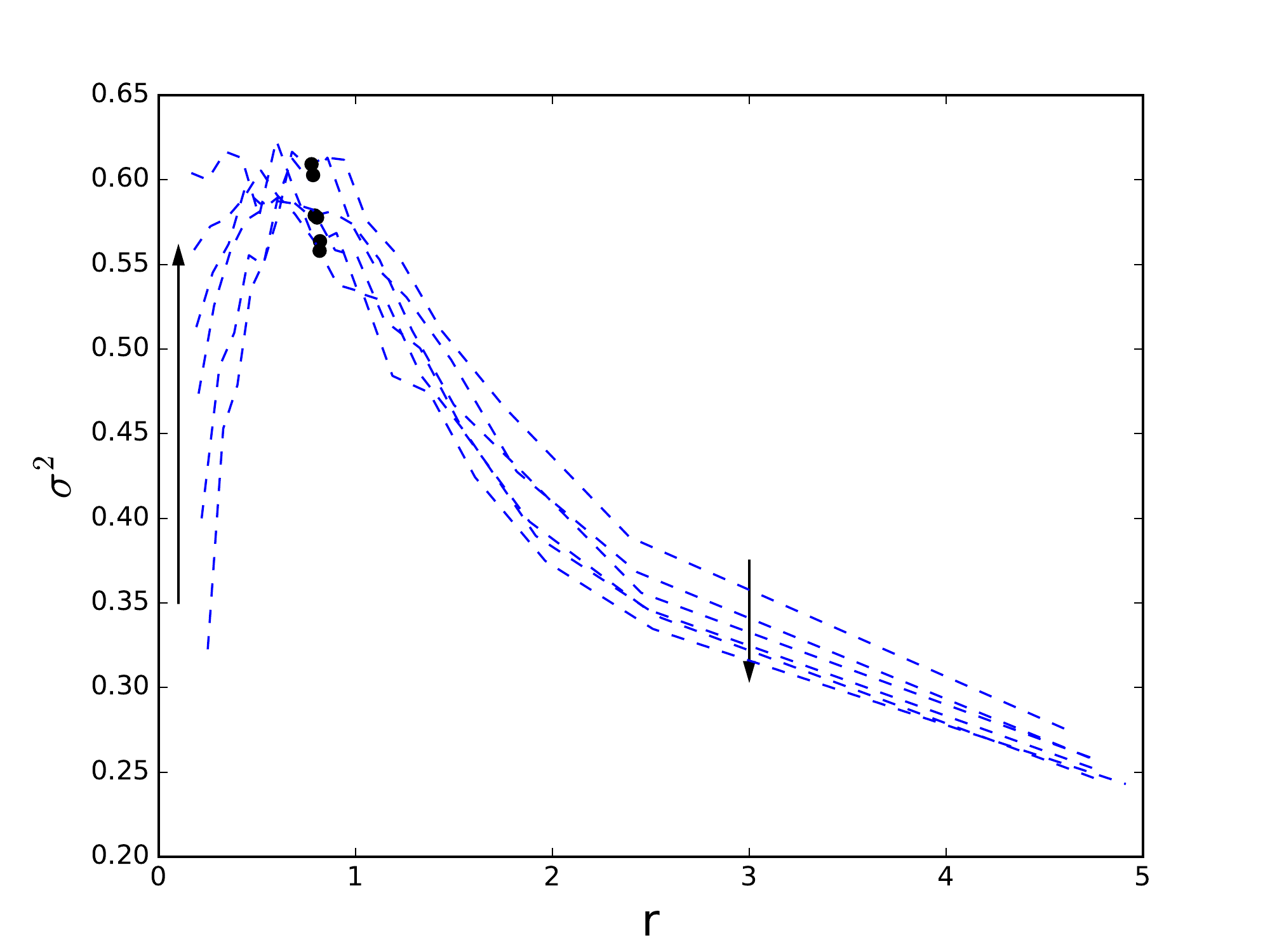} }} 
	\quad%
	\subfloat[]{{\includegraphics[trim=0.6cm 1.cm 1.4cm 0.9cm, width=\columnwidth]{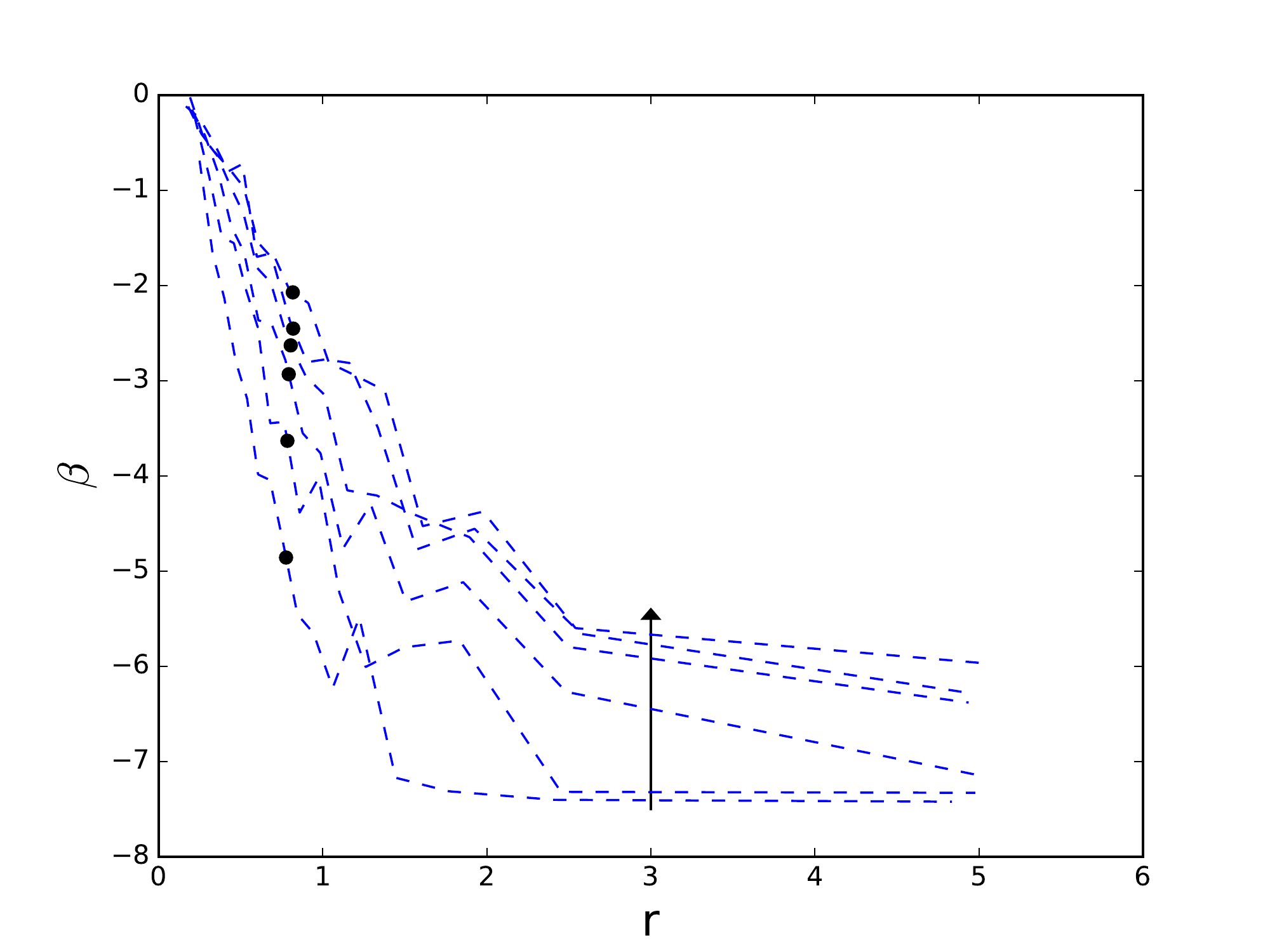} }} 
	\quad%
	\subfloat[]{{\includegraphics[trim=0.0cm 1.2cm 1.4cm 0.9cm, width=\columnwidth]{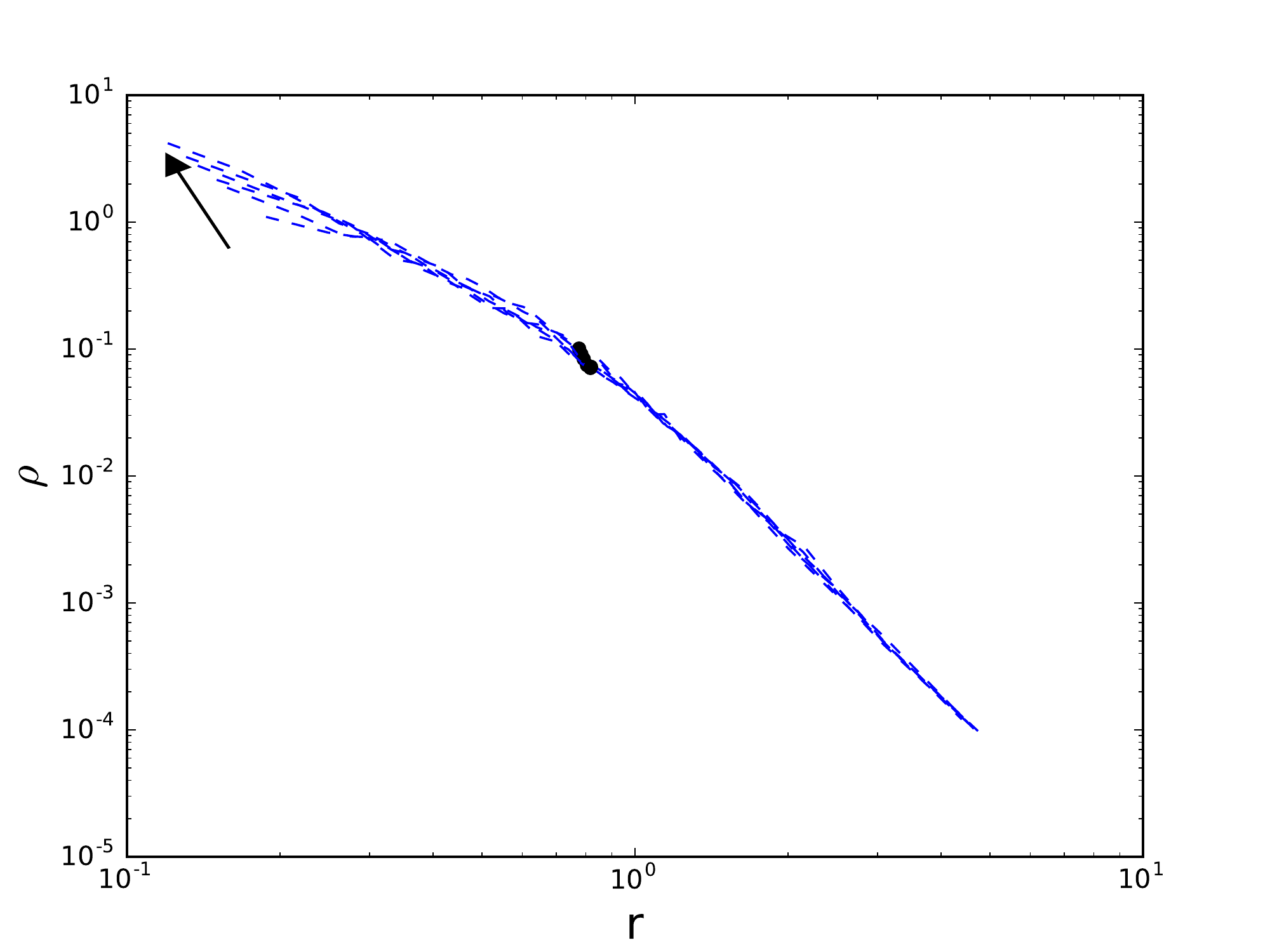} }}%
	\caption{\alv{Evolution of kinematics and structure of \dch{one realisation of} the model with $q = -16$.} Top:  velocity dispersion ($\sigma^2$) profile. Middle: anisotropy profile ($\beta$). Bottom: mass density profile ($\rho$). The black arrows show the direction of evolution \pgb{and the} $\bullet$ \pgb{marks the position of $r_h$.} The profiles are 20 H\'{e}non units apart and \alv{cover the range $[0,100]$ H\'{e}non units. For the construction of the $\beta$ and $\sigma^2$ profiles, the bins contain 500 particles, {but 250 particles} for the $\rho$ profiles.
}   The value of $\beta$ in the innermost bin at $t=0$ is $\approx -0.7$.}
\label{fig:prof}
\end{figure}

\alv{To support our interpretation, we have studied the evolution of the anisotropy in the central regions of the models by inspecting the behaviour of the value of $2\,T_r/T_\perp$ within the 10\% and 50\% Lagrangian radii (see Fig.~\ref{fig:main3}).  Tangentially anisotropic systems rapidly evolve so that the mass enclosed within the 10\% Lagrangian radius becomes isotropic {(within about 10\%)} in about 0.5 $t_{rh,i}$, and similarly within the half-mass radius, in about 5-6 $t_{rh,i}$.} 

 \alv{As for the initially radial model ($q=2$), while the mass within the 10\% Lagrangian radius becomes approximately isotropic in about 0.5-1 $t_{rh,i}$, the region within the half-mass radius remains radially anisotropic for longer (i.e., until about 10 $t_{rh,i}$); such a behaviour} is probably due to the contribution of high energy particles (on more radial orbits) passing within the half mass radius on their way to and from the halo. These particles would take a longer time to relax compared to the more bound particles. 


 \alv{To further characterise the initial phase of evolution of the anisotropic systems under consideration, we have also analysed the behaviour {of} the radial profiles of the velocity dispersion, anisotropy parameter $\beta$ and density; for brevity, in Fig.~\ref{fig:prof} we report exclusively the evolution of the model with $q=-16$. In this case, the initially negative gradient of the velocity dispersion is rapidly reduced and completely disappears in about 1 $t_{rh,i}$. Simultaneously, the amount of tangential anisotropy is significantly reduced, especially in the intermediate regions. Finally, the density profile in the outskirts of the model is virtually unchanged, \dch{though} 
     there is a \dch{noticeable steepening} 
     of the slope towards the centre \dch{in the first time interval}.} 



\section{Discussion}
\label{sec:Dis}

\subsection{Radial Orbit Instability}
\label{sec:roi}
\begin{figure}
		\includegraphics[trim=0.6cm 1.cm 1.4cm 0.7cm, width=1.\columnwidth]{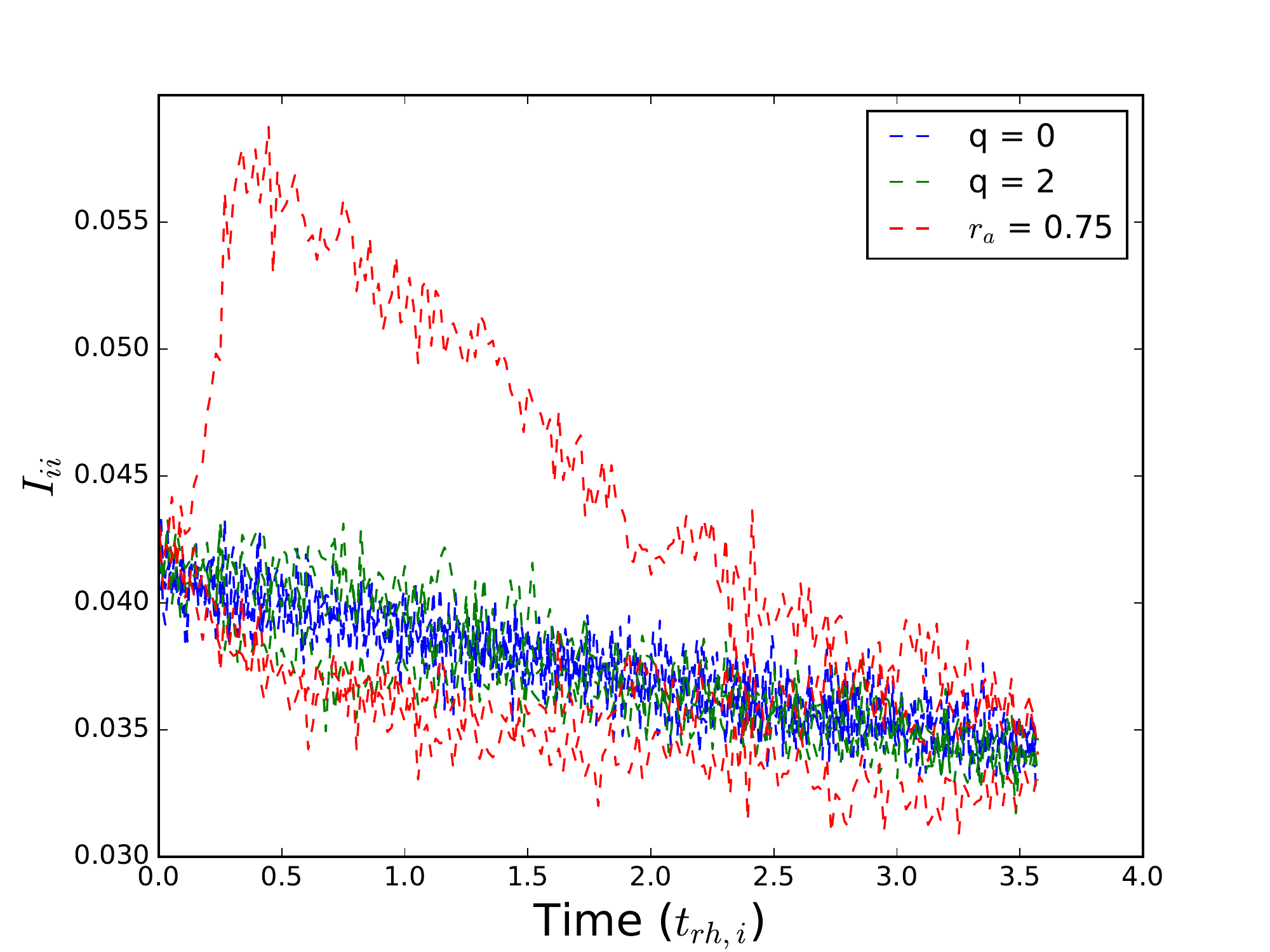}
                \caption{\alv{Evolution of the {second moments of the spatial distribution} 
                  } $I_{ii}$ ($\sum_j m_jx_i^2${, $i = 1,2,3$}) within the half mass radius of the Dejonghe $q= 0, 2$ \dch{models} and \dch{the} Osipkov-Merritt Plummer model with $r_a = 0.75$.  {The latter model provides the outlying curves at the top and bottom.} The $q = 2$ model is identical to the Osipkov-Merritt Plummer model with $r_a = 1.00$.  We find that the $q = 2$ models are stable and the $r_a = 0.75$ \dch{model is} 
                  unstable to radial orbit instability. }
    \label{fig:roi}
\end{figure}

\alv{By inspecting the behaviour of the radial modes and the quadrupole moment of equilibria evolved by means of a mean-field N-body code, \cite{Dejonghe1988} found that anisotropic {Osipkov-Merritt} Plummer models with $ r_a \lesssim 1.1 $\footnote{\dch{In this subsection the units of $r_a$ are those used in Sec.\ref{sec:OMmodels}, i.e. such that the scale radius of the Plummer density profile is unity.}} were unstable to the radial orbit instability.
  On the basis of our numerical simulations performed with a direct N-body code, we report that configurations with $r_a = 1.0$ (i.e. $q = 2$) are stable and \dch{those} with $r_a = 0.75$ are unstable, as visible from the time evolution of the {second moments of the spatial distribution} 
  illustrated in Fig.~\ref{fig:roi}.  }

\alv{We tend to attribute this slight variation in the value of the threshold for stability to the intrinsic differences between collisional and collisionless codes \dch{(see the last paragraph of this subsection)}, although it is worth mentioning that a number of subsequent investigations, performed on anisotropic configurations characterised by different density distributions, also reported revised stability thresholds. In particular, \alv{by monitoring the evolution of the axis ratio of models evolved with a `self-consistent field' N-body code}, \citet{MZ1997} found a stability limit of ${2T_r}/{T_\perp} = 2.31 \pm 0.27$  for selected configurations within the family of spherical anisotropic models proposed by \citet{Dehnen1993}.  {If we assume that the critical value is the same for other models,} this result implies that their stability boundary is below $ r_a = 1$ ($ q = 2 $) and above $r_a = 0.75$ (where ${2T_r}/{T_\perp} \approx 2.4$ for our models), which is consistent with our results.  Interestingly, a more recent numerical study performed \dch{by} \citet{TreBer2006} specifically examined the dependence of the occurrence of the radial orbit instability on the shape of the anisotropy profile in the central region of a stellar system; they identified stable states with  ${2T_r}/{T_\perp} \approx 2.75$, as emerging at the end of the process of cold collapse from homogeneous initial conditions.}

The core collapse time for the $r_a = 0.75$ model is $t_{cc}= 2140 \pm 30$, which is consistent with the $r_a = 1$ ($q = 2$) model in Table~\ref{tab:tab1} and in excess of the collapse time of the reference isotropic model.  From Fig.~\ref{fig:roi} we can see that the $r_a = 0.75$ model gradually returns to a spherical configuration within the half mass radius, after approximately 3 $t_{rh,i}$. \alv{We wish to note that} a detailed investigation of the effects of the radial orbit instability on the \dch{time} 
of core collapse is outside the scope of the present study. 

Since relaxation can affect the amount of anisotropy in significant way\dch{s}, this could change the stability boundary of \dch{the} radial orbit instability. As can be seen in Fig \ref{fig:main3}, in the collisional setting the more radial models initially evolve towards a more isotropic configuration. This  could cause a model which appears to be unstable when simulated with a collisionless code to become stable when simulated with a collisional code. It is perhaps the reason that we find the $r_a = 1.0$ model stable against the radial orbit instability\dch{, even though it has} 
previous\dch{ly} been found to be unstable (see \dch{above}
). This may imply a relaxation time dependency o\dch{f} the stability boundary. {A similar conclusion (i.e. that enhanced collisionality may tend to
suppress the radial orbit instability) was reached
by \cite{S2017} for anisotropic disk systems. }

\begin{figure}
	\includegraphics[trim=0.6cm 0.9cm 1.4cm 0.8cm, width=\columnwidth]{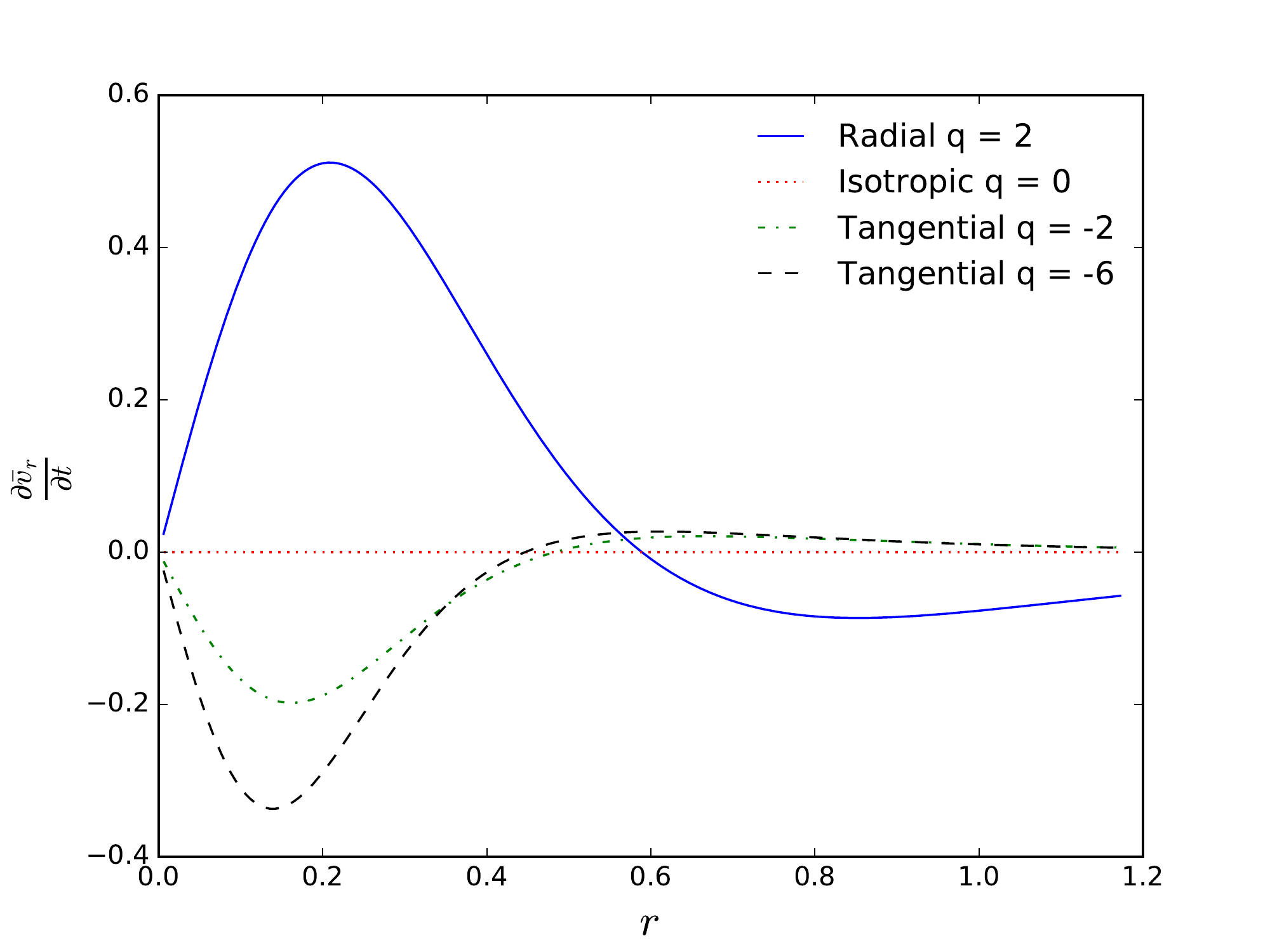}
        \caption{The predicted response of \alv{selected} anisotropic Plummer models \pgb{(in H\'{e}non units), see Section~\ref{sec:jeans} for details.}}
    \label{fig:jeans}
\end{figure}

\subsection{ A theoretical interpretation based on hydrostatic equations }
\label{sec:jeans}

 \alv{We attempt to provide an interpretation of} the early evolution of the anisotropic Plummer models by using the radial Jeans equation. Assuming spherical symmetry, the radial Jeans equation in spherical coordinates can be written as
\begin{equation}
\frac{\partial \bar{v}_r}{\partial t} = - \frac{1}{\rho}\frac{\partial (\rho \sigma_r^2)}{\partial r} - \frac{(\dch{2}\sigma_r^2-\sigma_t^2)}{r} - \frac{\partial\Phi}{\partial r}. \label{eq:jeans}
\end{equation}
Under some simplifying assumptions we can predict how the system would \alv{respond} as it relaxes towards an isotropic velocity distribution. \alv{Specifically, we assume} that $\rho$ and $\sigma_{tot}^2$ (where $\sigma_{tot}^2 = \sigma_{r}^2+\sigma_{t}^2$) remain constant and that, after \alv{one} central relaxation time $t_r(0)$, $\sigma_r^2$ becomes $\sigma_r^2 +  \frac{t_r(0)}{t_r(r)}(\frac{1}{3}\sigma_{tot}^2 - \sigma_{r}^2)$ and $\sigma_t^2$ is simply $ \sigma_{t}^2 =   \sigma_{tot}^2 - \sigma_{r}^2$.

Using this result we can evaluate the right-hand side of the radial Jeans equation{, Eq.~(\ref{eq:jeans}).} 
We have \alv{performed this calculation numerically for different values} of $q$ and plotted the results in Fig~\ref{fig:jeans}. This theory predicts that \alv{the central regions of a stellar system will experience an acceleration; the sign of such an acceleration term depends on the flavour of the anisotropy, with radially-biased anisotropic \dch{models} experiencing a positive (outward) acceleration and tangentially-biased anisotropic systems a negative (inward) one. Such a behaviour is consistent with the accelerated (delayed) early phase of evolution characterising the tangential (radial) models which we have considered in Section~\ref{sec:results}. }

\subsection{ A theoretical interpretation based on a fluid model of two-body relaxation}\label{sec:theoryII}

{The foregoing theoretical interpretation fails for the isotropic
model (which contracts from $t = 0$).  Also, as stated in the assumptions, it is based on the idea
that anisotropy evolves without any change in $\rho$ and
$\sigma_{tot}$; this gives rise to a significant departure from
hydrostatic balance, and a significant acceleration.  In fact,
$\rho,\sigma_{tot}$ and $\sigma_r$ all change simultaneously, leaving a
small hydrostatic imbalance which gives rise to a mean flow on a time
scale of the relaxation time.  The following theory recognises this
simultaneous evolution, and predicts the initial rate of evolution of
central quantities, including the isotropic case.  It does not,
however, say anything about the radial profile of this evolution,
which would require a full numerical solution of the equations we are
about to discuss.}

{\citet{LS1991} describe a gas model of a spherical,
anisotropic star cluster, in which the evolution is described in terms
of the density $\rho$, and radial and transverse velocity dispersions
$\sigma_r^2, \sigma_t^2$ (or
equivalent variables), which are functions of radius $r$ and time
$t$.  Relaxation drives fluxes of mass and of radial and transverse kinetic
energy, which are described by three radial velocities $u,v_r$ and
$v_t$.  
While \citet{LS1991} describe examples of the evolution of the full spatial structure of a system, in the present application we confine \alv{our} attention to the initial rate of change of one of the central properties, which can be determined without any numerical integration of the partial differential equations of the model.}

{The central property on which we focus is the local specific entropy.  Details are given in Appendix \ref{sec:centralentropy}, where it is shown that, for initial conditions of the anisotropic Plummer models studied in the present paper, its initial rate of change is given by the equation
\begin{equation}
  \frac{d}{dt}\ln\left(\frac{\sigma^3}{\rho}\right)
  = -\frac{10\lambda(3+2q)\sqrt{6-q}}{\sqrt{\pi}}\frac{\log N}{N}\left(\frac{GM}{a^3}\right)^{1/2},\label{eq:entropy}
\end{equation}
where $a$ is the scale radius of the Plummer model.
Incidentally,
exactly the same result (except for a change to the
dimensionless coefficient) could be obtained from the isotropic model
of \citet{lydenbell1980}.  As those authors mention, the
expression on the left of Equation (\ref{eq:entropy}) is essentially the
rate of change of the specific entropy, $s$, except for its sign.}

{For an isotropic Plummer model, we set $q=0$, and see that $\dot s
<0$.  Thus either $\sigma$ decreases or $\rho$ increases initially.  $N$-body
models show that an isotropic Plummer model actually does both
initially, though after a small adjustment $\sigma$ begins to
increase.  For anisotropic models there are two differences to
notice.  First, the initial rate of change of $s$ changes to $\dot
s>0$ for $q < -3/2$.  \dch{(Note that this happens to be the value at which a temperature inversion near the centre of the model first appears (as $q$ decreases); see Sec.\ref{sec:einstein}.)}  Thus, either $\sigma$ increases or $\rho$
decreases, or both.  Second, for large negative $q$ (strongly
tangentially anisotropic initial models) the time scale of the initial
evolution of $s$ becomes very short.  The reason for this is the small
central velocity dispersion, which leads to a short relaxation time.}

\begin{figure}
  \includegraphics[trim=0.8cm 1.2cm 1.4cm 0.8cm, width=\columnwidth]{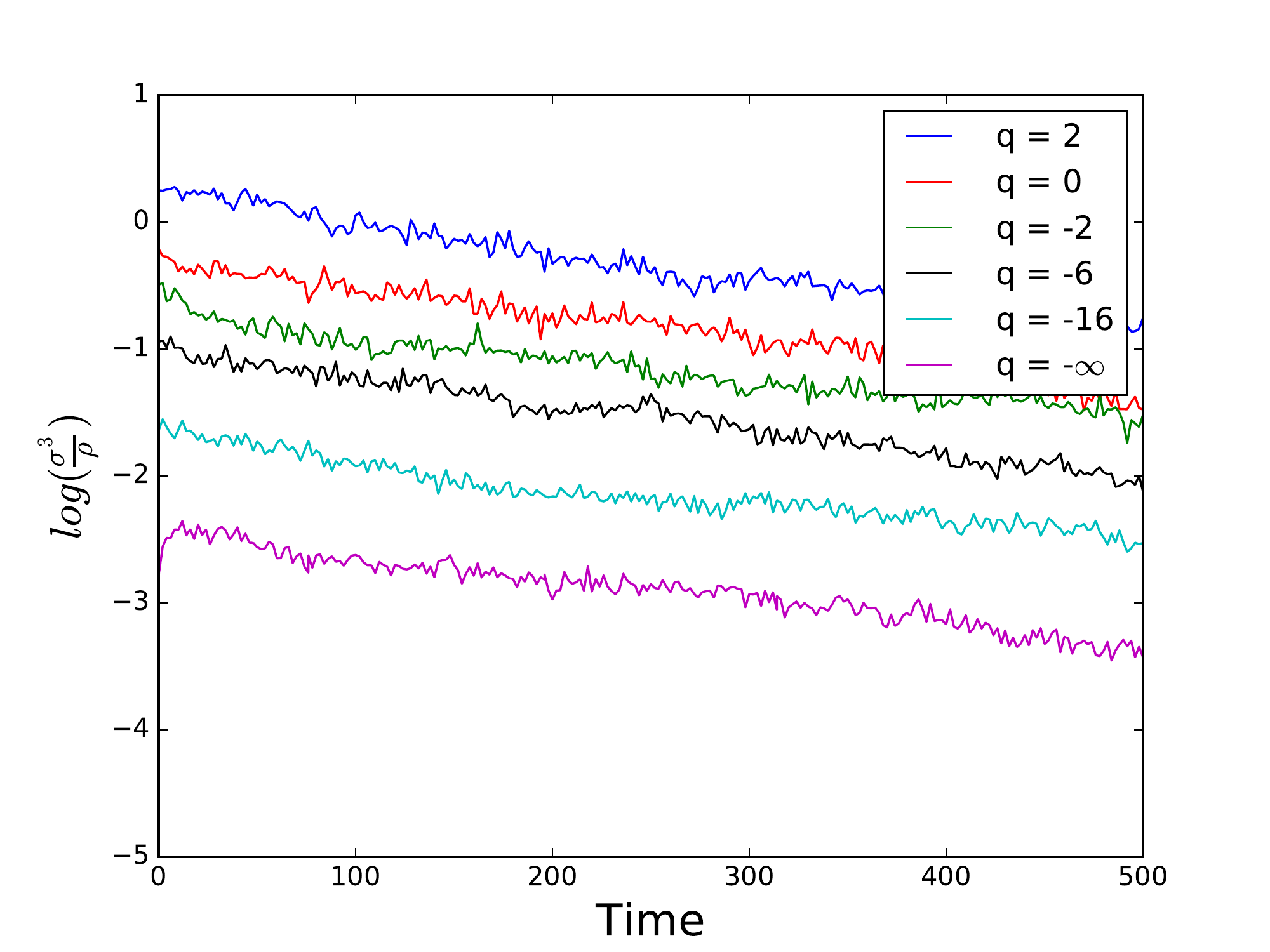}
  \caption{Early evolution of the central specific entropy in the \alv{anisotropic Plummer models illustrated in Fig.~\ref{fig:main}}.  {The initial value of the specific entropy is an increasing function of q} (\pgb{ one realization is shown for each value of q}).  The velocity dispersion and density are computed from the 500 innermost particles. \alv{Time is expressed in H\'{e}non units.}}
  \label{fig:s500}
\end{figure}

{\alv{The evidence based on our $N$-body simulations is illustrated in Fig.~\ref{fig:s500}}.  Two points at least are clear, despite the sampling fluctuations.  The first is that the most extremely tangentially anisotropic model ($q=-\infty$) exhibits the most rapid early evolution, and the second is that its initial rate of change is positive.  While these observations are consistent with the predictions we have made on the basis of Equation~(\ref{eq:entropy}), we readily concede that there is no sign that the transition from positive to negative rate of change takes place near $q = - 3/2$, as predicted by the model.}

{It might be possible to improve this theory by computing the initial rate of change of the central value of the density function $f(\br,\bv)$, which is also closely related to the central specific entropy.  The rate of change can, in principle, be evaluated using an anisotropic form of the Fokker-Planck equation, along the lines of the isotropic calculations presented in \citet[][section~7]{1988MNRAS.230..223H}.}

\subsection{Some considerations on the nature of the phenomenon}

One may wonder if the behaviour \dch{in models with extreme tangential anisotropy} 
is an instability in \alv{the} same sense as, for example, the gravothermal instability. Mathematically, an instability in a dynamical system is a property of an equilibrium. While the models considered here are equilibria in the collisionless sense (i.e. they satisfy the collisionless Boltzmann equation), they are not collisional equilibria. For collisional systems the only true equilibria {of finite mass} are highly artificial systems contained within a perfectly reflecting wall \citep[e.g.][]{Antonov1962}. \alv{We note that} the models commonly used in stellar dynamics are not collisional equilibria, therefore, strictly speaking, the term instability does not apply to any behaviour associated with core collapse, \alv{recognised as a collision-induced phenomenon}. However, since in the case of collisional systems contained within a spherical boundary (which do have equilibria), the gravothermal instability is a \alv{genuine} instability, and since it is fundamentally the same \alv{physical} mechanism which drives core collapse also in models without strictly defined equilibria, the terminology seems \alv{none the less} appropriate. Since there does not appear to be an artificial case where the highly tangential models are collisional equilibria (as is the case for gravothermal instability), describing this behaviour as an instability does not seem appropriate

\dch{One may still ask the question whether the phenomenon of rapid early collapse in models with extreme tangentially biased anisotropy is collisional (as we assumed in Sections \ref{sec:jeans} and \ref{sec:theoryII}) or collisionless.  To clarify this we carried out one simulation for the case $q=-\infty$ which was designed exactly as in Sec.\ref{sec:numerical} except that $N = 32768$.  Up to time at least 100 H\'enon units in the larger model, the evolution of the Lagrangian radii, up to at least the 10\% mass fraction, is almost indistinguishable (within fluctuations) from that in our typical 8k model {\sl when the time axis is scaled by the initial relaxation time}.  This leaves no room for doubt that the very rapid early collapse is indeed a collisional phenomenon, though it does not prove that the system is stable in the collisionless sense.  Unfortunately the literature on collisionless stability of systems with circular orbits does not seem to be conclusive (for the Dejonghe model with $q=-\infty$).  The uniform sphere is stable \citep{MFE1971,Fridman1984}, which might have been thought relevant to dynamics close to the centre of a model with a Plummer density distribution.  On the other hand \cite{Weinberg1991}, drawing partly on the discussion in \cite{Poly1987}, mentions an expectation that any model populated by nearly circular orbits should be unstable, at least in extreme or pathological cases.  Whatever the theoretical expectation, however, there is no sign of collisionless instability in our extreme models, if only because it is swamped by collisional behaviour taking place on the the very small central two-body relaxation time.}


\subsection{\alv{Application to other anisotropic models}}
\label{sec:geniso}
\alv{The dynamical response of an anisotropic stellar system, especially the time scale of its evolution towards a condition of isotropy in the velocity space, depends in principle on the spatial distribution of the primordial anisotropy.  In the present study, we have intentionally limited our investigation to the case of configurations characterised by a Plummer density distribution, generalised by \citet{Dejonghe1987} to allow anisotropy in the velocity space. We have chosen such a family because of the simplicity of the radial profile of the anisotropy parameter (see Equation~\ref{eqn:beta}), which, at the half-mass radius, is such that $\beta (r_h) \dch{= 2^{-5/3}q}
  $. }

\alv{We emphasise that the early phase of the ``anisotropic response" has direct implications on the \dch{time} 
  of core collapse\dch{: in cases of high tangential anisotropy, for instance, it quickly takes the model a long way towards core collapse (Fig.\ref{fig:main2})}. Therefore, the detailed properties of the relationship between the \dch{time} 
  of core collapse and the initial strength of the primordial anisotropy (see Fig.~\ref{fig:tcc}) strictly depend on the initial structure of the three-dimensional velocity space. None the less, on the basis of the theoretical interpretations provided in Sections~ \ref{sec:jeans} and \ref{sec:theoryII}, we expect the general behaviour (i.e., acceleration or delay in the presence of tangential or radial primordial anisotropy, respectively) to apply to a wide range of spherically symmetric, anisotropic equilibria. The interpretation based on hydrostatic equations presented in Section~\ref{sec:jeans} could easily be repeated for a different set of \alv{anisotropic models}. 
}



\begin{figure}
	\includegraphics[trim=0.7cm 1.2cm 1.6cm 1cm, width=\columnwidth]{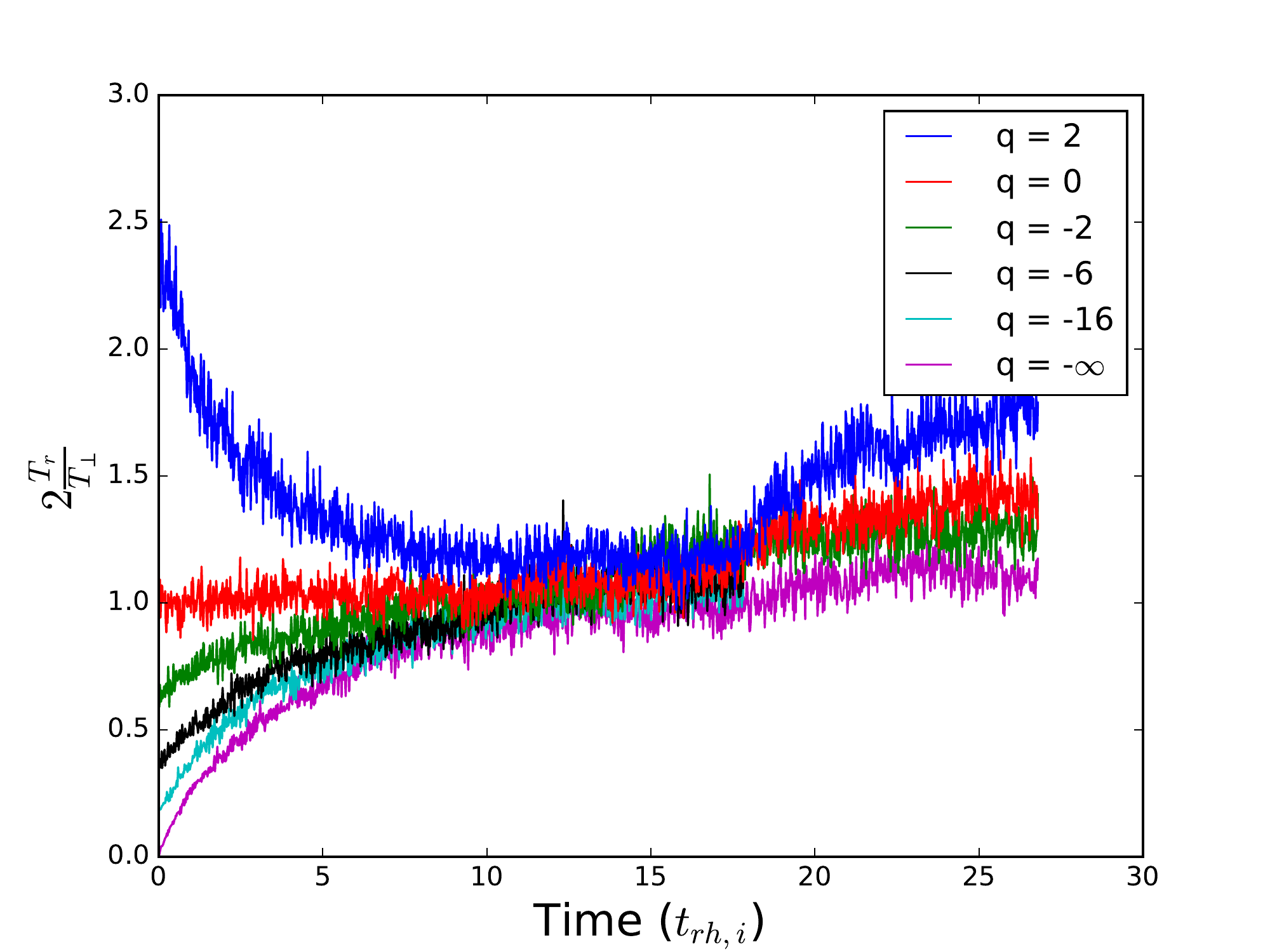}
    \caption{\alv{Exploration of the transition between primordial and evolutionary anisotropy,} { with plots of the} time evolution of $2\,{T_r}/{T_\perp}$ in a shell around $r_h$ for the models illustrated in Fig.~\ref{fig:main} (\pgb{One realization is shown for each value of q}). All models eventually become radially anisotropic, \alv{to a degree which depends on the flavour and strength of the primordial anisotropy (see Section~\ref{sec:dyanio} for details).}}
    \label{fig:trtplagr}
\end{figure}

\subsection{Primordial vs. evolutionary anisotropy }
\label{sec:dyanio}
A number of studies have noted that initially isotropic models \alv{develop in their outer regions a significant degree of anisotropy of the velocity space, especially after core collapse, \dch{whether} 
  in isolation or in the presence of a tidal field \citep[e.g., see][]{GH1994, Zocchi2016, TVV2016}}. This is usually attributed to 
dynamical scatter\dch{ing} near the time of \dch{core} collapse 
and interactions with binary stars afterwards. 

\alv{In order to explore the transition from the regime of primordial {anisotropy to that of} 
  evolutionary anisotropy, we have measured the indicator $2\,{T_r}/{T_\perp}$ in a shell around the location of the half-mass radius, as depicted in Fig~\ref{fig:trtplagr}. All models eventually develop radial anisotropy, to a degree which depends on the flavour and strength of the primordial \dch{anisotropy} 
  (i.e., the model developing the highest [lowest] value of evolutionary radial anisotropy is the one characterised by the strongest primordial radial [tangential] anisotropy).  {Though it might seem obvious that a model with greater primordial radial anisotropy will show greater radial anisotropy at later times, it is not quite that simple.}  
  {Since}  the tangentially-biased models 
  undergo core collapse earlier in their evolution, {they} could, in principle,  have had more time to build up the radial {anisotropy.} 
  In fact, we believe that the amount of post-core-collapse anisotropy in the intermediate and outer regions of the models should be interpreted as the result of the cooperation or competition between the growth of the genuine evolutionary radial anisotropy and the permanence of the primordial \dch{anisotropy}
, which may still be preserved in the outer portions of the system, as they are characterised by a longer relaxation time. In any case, we argue that the physical origin of the evolutionary radial anisotropy is {qualitatively} the same as in the isotropic case (i.e., ejection of stars from the core to the halo preferentially on radial orbits).  } 


\section{Conclusions}
\label{sec:sum}

We have studied the effect of primordial anisotropy in a series of \dch{equilibrium} models \alv{characterised by the same spatial distribution of mass, defined as in the spherical isotropic Plummer model, \dch{the same initial half-mass relaxation time, but varying anisotropy.}   The main result of our investigation is that there is a clear dependence of the core collapse time on the amount of velocity anisotropy of the stellar system. Specifically, we report that models characterised by radial anisotropy require a longer time to reach core collapse, compared to the reference isotropic case; tangentially anisotropic models require a shorter time, with the most tangential case (Einstein Sphere) being the absolute fastest to collapse.} \alv{On the basis of the series of N-body simulations we have performed, we therefore conclude that, in the presence of the same initial structural properties, {there} 
  appears to exist a {monotonic relationship} between the amount of primordial anisotropy and the {time} 
  of core collapse for spherical, isolated collisional systems.}
\dch{The effect is large,
the core collapse time for the most tangentially anisotropic models being three times smaller than for the most radially anisotropic models we studied, in which the ratio of  the kinetic energies in radial and transverse motions is approximately twice the value in the standard (isotropic) Plummer model (Fig.\ref{fig:tcc}).}

\alv{We interpret this behaviour as resulting from the early phase of collisional evolution during which the initially anisotropic configurations rapidly evolve towards a condition of isotropy in the velocity space. From a numerical {study} 
  of the radial component of the Jeans equations (under some simplifying assumptions) we expect the inner parts of {a tangentially anisotropic} 
  system to lose support from the tangentially-biased motions as the system relaxes towards an isotropic velocity distribution.}  For the case of radial models we predict the opposite effect, therefore the systems tend to expand as they evolve towards isotropy.
For the case of models which are {\sl highly} tangentially anisotropic (e.g. Einstein Sphere), the initial evolution involves a catastrophic collapse, {because of the very short central relaxation time.}  
Once the models become isotropic in the central regions, they all evolve along a similar evolutionary track, \alv{which may be identified also in the energy space.} 


\section*{Acknowledgements}

{ We thank an anonymous referee for several comments including
pertinent references.} \alv{We are grateful to Mark Gieles, Ralf Klessen, Rainer Spurzem, Enrico Vesperini, and Alice Zocchi for interesting conversations about the role of anisotropy in the dynamical evolution of star clusters.} \alv{All authors} acknowledge support from the Leverhulme Trust \alv{(Research Project Grant, RPG-2015-408)}, and ALV also from the EU Horizon 2020 program \alv{(Marie Sklodowska-Curie Fellowship, MSCA-IF-EF-RI 658088)}. 









\appendix

\section{{Evolution described by an anisotropic gaseous model of two-body relaxation}}\label{sec:centralentropy}

{In this Appendix we describe details of the derivation of an equation which is the focus of Section~\ref{sec:theoryII}.  As mentioned there the equation is based on a model for the evolution of an anisotropic spherical system under the action of two-body relaxation, which drives fluxes of radial and transverse kinetic energy with velocities $v_r,v_t,$ respectively.  The model is due to \citet{LS1991}
. Actually, these authors describe two
variants of the model, one of which (their Model B) would lead to
$v_r$ and $v_t$ differing by a term of order $r$ at small radii in an
anisotropic Plummer model.  As those authors note, such behaviour is
incompatible with a finite central density, and so we adopt their
Model A, in which $v_r = v_t = v$, where $v$ is the velocity of
transport of kinetic energy.}

{Much as in the isotropic gas model of \citet{lydenbell1980}, the
Lagrangian fluxes are taken to be proportional to temperature
gradients, and Louis \& Spurzem (see their equation 23) take
\begin{equation}
  v - u + \frac{\lambda}{4\pi G\rho
    T}\frac{\partial\sigma^2}{\partial r} = 0\label{eq:v}
\end{equation}
where $\lambda$ is a dimensionless constant,
\begin{equation}
\sigma^2 = (\sigma_r^2 + \sigma_t^2)/3, \label{eq:sigma}
\end{equation}
and $T$ is a relaxation time defined by
\begin{equation}
  T = \frac{9\sigma^3}{16\sqrt{\pi}G^2m\rho\log N},\label{eq:T}
\end{equation}
in which $m$ is the stellar mass.}  \dch{ Note also that, in this account of the model, we have adopted from the present paper the definition of $\sigma_t^2$ as the mean square two-dimensional transverse velocity.}

{In the anisotropic Plummer model, the velocity dispersions are given
by \citet[][equations 22a, b]{Dejonghe1987}, whence we find that, at
small radius, we have
\begin{eqnarray}
  \sigma_r^2 &=&
  \frac{1}{6-q}\frac{GM}{R}\left(1-\frac{1}{2}\frac{r^2}{R^2}\right) +
  O(r^4)\label{eq:sigmar}\\ \sigma_t^2 &=&
  \frac{2}{6-q}\frac{GM}{R}\left(1-\frac{1}{2}(q+1)\frac{r^2}{R^2}\right) +
  O(r^4),\label{eq:sigmat}
\end{eqnarray}
though we have reinstated $G$, the cluster mass ($M$) and the scale
radius of the Plummer model ($R$), which Dejonghe scaled away.  \dch{(We avoid use of the more common $a$, which is reserved for anisotropy in the model, and is mentioned below.)}  Hence
we find from Equations (\ref{eq:v}), (\ref{eq:sigma}) and (\ref{eq:T}) that
\begin{equation}
  v - u = \frac{\lambda(3+2q)}{9(6-q)}\frac{r}{T} + O(r^3)\label{eq:vmu}
\end{equation}
where we have made use of a standard expression for the value of
$\rho$ at the centre of a Plummer model.}

{In order to evaluate the effect of these fluxes on central quantities,
we substitute them into two of the four moment equations of the model
\citep[][equations 43-46]{LS1991}.  The first of these is
mass continuity, i.e.
\begin{equation}
  \frac{\partial\rho}{\partial t} +
  \frac{1}{r^2}\frac{\partial}{\partial r}(\rho u r^2) = 0.
\end{equation}
We suppose that $u = u'r + O(r^3)$, where $u'$ is its central
$r$-derivative, and so find the obvious result that
\begin{equation}
  \dot\rho + 3\rho u' = 0\label{eq:rhodot}
\end{equation}
to lowest order in $r$, and the variables $\dot\rho, \rho$ also denote
central values.}

{Next comes the hydrostatic equation.  This is equivalent to our
Equation (\ref{eq:jeans}), except that the acceleration is absent.  In this form it is
satisfied by the anisotropic Plummer model, and it gives no
information on the rate of change of any central quantity.}

{The last two moment equations are flux balance equations for $p$ and
the product $pa$, where
\begin{equation}
p = \rho\sigma^2\label{eq:pdef}
\end{equation}
is the pressure and $a$ is
a measure of the anisotropy.  Only the first equation is useful, as $a
= 0$ to order $r$ near the centre.  The first equation, however, is
easily handled like the continuity equation, giving
\begin{equation}
  \dot p + 5pv' = 0\label{eq:pdot}
\end{equation}
to lowest order in $r$.}

{A suitable linear combination of Equations (\ref{eq:rhodot}) and
(\ref{eq:pdot}), with use of Equation (\ref{eq:pdef}), gives
\begin{equation}
  \frac{d}{dt}\ln\left(\frac{\sigma^3}{\rho}\right)
  +\frac{15}{2}(v'-u') = 0,
\end{equation}
and so with Equations (\ref{eq:vmu}) and (\ref{eq:T}) we find that
\begin{equation}
  \frac{d}{dt}\ln\left(\frac{\sigma^3}{\rho}\right)
  = -\frac{10\lambda(3+2q)\sqrt{6-q}}{\sqrt{\pi}}\frac{\log N}{N}\left(\frac{GM}{R^3}\right)^{1/2},
\end{equation}
where we have \dch{again} used \dch{the} 
standard expression for the central density
in a Plummer model, and Equations (\ref{eq:sigma}), (\ref{eq:sigmar}) and
(\ref{eq:sigmat}) for the central value of $\sigma$.  This is the equation which\dch{, with one change of notation,} is discussed in Section \ref{sec:theoryII}.}


\bsp	
\label{lastpage}
\end{document}